\documentclass[english,superscriptaddress,floatfix,twocolumn,oneside,pra]{revtex4-1}
\usepackage[english]{babel}
\usepackage[utf8x]{inputenc}
\usepackage{microtype}
\usepackage{graphicx}
\usepackage{amsmath,amsfonts}
\usepackage{siunitx}
\usepackage[colorlinks=true,linkcolor=black,citecolor=black]{hyperref}
\usepackage{bm}
\usepackage{bbold}
\usepackage{natbib}
\usepackage{csquotes}

\usepackage[draft]{fixme}

\usepackage[capitalise]{cleveref}

\renewcommand\vec[1]{\ensuremath\boldsymbol{#1}}

\newcommand{\unitm}{\ensuremath{\mathbb{1}}}

\crefname{section}{Sec.}{Sec.}

\newcommand{\affA}{Department of Physics and Astronomy, Aarhus University, DK-8000 Aarhus C, Denmark}
\newcommand{\affB}{Aarhus Institute of Advanced Studies, Aarhus University, DK-8000 Aarhus C, Denmark}

\date{\today}
\begin{document}

\title{A controllable two-qubit swapping gate using superconducting circuits}

\author{S. E. Rasmussen}
\email{stig@phys.au.dk}
\affiliation{\affA}
\author{K. S. Christensen}
\affiliation{\affA}
\author{N. T. Zinner}
\email{zinner@phys.au.dk}
\affiliation{\affA}
\affiliation{\affB}

\begin{abstract}
	In this paper we investigate a linear chain of qubits and determine that it can be configured into a conditional two-qubit swapping gate, where the first and last qubits of the chain are the swapped qubits, and the remaining middle ancilla qubits are controlling the state of the gate. The swapping gate introduces different phases on the final states depending on the initial states.
	In particular we focus on a chain of four qubits and show the swapping gate it implements. We simulate the chain with realistic parameters, and decoherence noise and leakage to higher excited states, and find an average fidelity of around 0.99.
	We propose a superconducting circuit which implements this chain of qubits and present a circuit design of the circuit. We also discuss how to operate the superconducting circuit such that the state of the gate can be controlled. Lastly, we discuss how the circuit can be straightforwardly altered and may be used to simulate Hamiltonians with non-trivial topological properties.
\end{abstract}

\maketitle

\section{Introduction}\label{sec:intro}

A universal set of quantum gates can consist entirely of two-qubit gates \cite{DiVincenzo1995}. If a quantum information processor is to be created, it is therefore desirable to have a number of two-qubit gates which can be implemented without too much difficulty. One of the most promising candidates for the base of such a processor is superconducting qubits, where single-qubit gate operations are performed with gate fidelities well above 0.99 \cite{Buluta2011,Gustavson2013,Reagor2018,Rol2017,Sheldon2016,Chen2016,Barends2014}, which is the lower bound for performing fault-tolerant quantum computing, using error correction surface codes \cite{Raussendorf2007,Corcoles2015,OBrien2017,Fowler2014}. However, the fidelity of two-qubit gates are still trailing behind. In 2011, IBM demonstrated a fixed coupling gate with a fidelity up to 0.81 \cite{Chow2011,Li2008,Rigetti2010,Paraoanu2006}, while fidelities up to 0.994 have been reported in 2014 in a controlled phase-gate \cite{Barends2014,Kelly2014,Chen2014}, and in 2016, IBM achieved a fidelity of 0.991 in the cross-resonance gate \cite{Sheldon2016a}. Other notable two-qubit gates that have performed with a fidelity of above 0.9 are: The $i$\textsf{SWAP} and $\sqrt{i\mathsf{SWAP}}$ gates \cite{Chen2016,McKay2016,Dewes2012,Salathe2015}, the $b$\textsf{SWAP} gate \cite{Poletto2012}, and the resonator induced phase gate \cite{Paik2016}.

In this paper, we investigate what kind of quantum mechanical two-qubit gates a linear chain of qubits implements. We further propose a way of implementing such a chain using superconducting qubits. We show that such a chain, with an average fidelity around 0.99, swaps the end qubits which receive a phase depending on the configuration of the linear chain. The swapping operation is controlled on the middle qubits, acting as ancilla qubits. All in all this implements a conditional two-qubit swapping gate.

This paper is organized as follows: In \cref{sec:H} we introduce the Hamiltonian of the system, and the requirements to it. This is followed by \cref{sec:swappingGate} where we present the swapping gate which the Hamiltonian implements, and perform a numerical investigation of the average fidelity of the gate when varying the parameters of the system. Then, in \cref{sec:SCC}, we present a superconducting circuit which implements the desired Hamiltonian in the case of four qubits. We also present a chip design of the circuit. We discuss the effect of leakage and decoherence noise in a realistic implemented system via a numerical simulation, using realistic parameters related to the circuit, in \cref{sec:leakage}. In \cref{sec:circuitOutlook} we discuss how to mend the superconducting circuit into simulating other quantum systems, thus showing the utility of the circuit. Finally in \cref{sec:conclusion} we summarize and conclude the paper.

\section{The system}\label{sec:system}

We claim that by using a linear Heisenberg model we can implement a two qubit swapping gate. We start by presenting the Hamiltonian of the system, and then explains how it yields the gate.

\subsection{The Hamiltonian}\label{sec:H}

The Heisenberg model has many interesting applications on its own, from the study of quantum phase transitions \cite{Vojta2003,Hu1984} and magnetism \cite{Morita2018} to exploring topological states such as spin liquid states \cite{Depenbrock2012}. It is also closely related to the Hubbard model. 
Here we consider a linear Heisenberg spin chain consisting of $N$ spins (or qubits). In the Schr\"{o}dinger picture the linear Heisenberg spin model takes the form

\begin{align}
H =&-\frac{1}{2}\sum_{j=1}^N \Omega_j \sigma^z_j\\ &+  \sum_{j=1}^{N-1} \left[J^x_{j} (\sigma^x_j\sigma_{j+1}^x + \sigma^y_j\sigma_{j+1}^y)  + J^z_{j} \sigma^z_j \sigma^z_{j+1} \right], \nonumber
\end{align}
where $\sigma_j^{x,y,z}$ are the Pauli spin matrices, $\Omega_j$ denotes the frequency of qubit $j$, and the $J^{x,z}_{j}$'s denotes the coupling between the $j$'th and $(j+1)$-th qubit. This means that we consider only nearest neighbor $XXZ$ interactions. Note that we use $\hbar = 2e = 1$ throughout this paper.

We now follow Ref. \cite{Marchukov2016} and assume a spatially symmetric spin chain, meaning that $\Omega_j = \Omega_{N+1-j}$ and $J^{x,z}_j = J^{x,z}_{N-j}$. In order to study the role of the interactions we transform into the interaction picture choosing the noninteracting Hamiltonian as
\begin{equation}
	H_0 = -\frac{1}{2}\Omega_1\sum_{j=1}^N  \sigma^z_j,
\end{equation}
which yields the interaction Hamiltonian 
\begin{align}\label{eq:HI}
H_I =&-\frac{1}{2}\sum_{j=2}^{N-1} \Delta_j \sigma^z_j\\ &+  \sum_{j=1}^{N-1} \left[J^x_{j} (\sigma^x_j\sigma_{j+1}^x + \sigma^y_j\sigma_{j+1}^y)  + J^z_{j} \sigma^z_j \sigma^z_{j+1} \right], \nonumber
\end{align}
where the detuning is $\Delta_j = \Omega_j-\Omega_1$ and we have used the rotating wave approximation to neglect interaction terms which obtain a time-dependent phase of $e^{\pm 2i\Omega_1 t}$. This is justified under the assumption that $\Omega_1 \gg J_j$, which we assume for the rest of the paper.

Although the result of Ref. \cite{Marchukov2016} is valid for any $N\geq 4$ we will now focus on the case of $N=4$. This is partly because it simplifies the arguments while the ideas remain intact, and partly because a physical implementation, as discussed in \cref{sec:realization}, is more easily done with fewer qubits. See the Supplementary Material \cite{supplMat} for a discussion of the case of five qubits.
With only four qubits, we are left with just one detuning, why we drop the subscript, $\Delta \equiv \Delta_2$, and four interaction terms, $J_{1,2}^{x,z}$.
The last requirements for the gate relates these parameters; the first is $\Delta = \Delta_{\pm} \equiv 2(J_2^z \pm J_2^x)$, in accordance with Ref. \cite{Marchukov2016}, while the second requirement is $J_1 \equiv J_{1}^x = J_1^z$. For a derivation of these requirements, see the Supplementary Material \cite{supplMat}. A schematic model of the system is seen in \cref{fig:SCCmodel}(a).

\subsection{The two-qubit swapping gate}\label{sec:swappingGate}

We claim that the above Hamiltonian, consisting of four qubits, implements a two-qubit swapping gate, where the first and the last qubits are the swapped qubits, while the middle ancilla qubits control the state of the gate. We thus have a multi-qubit controlled gate, where the combined state of the control qubits determine the state of the gate, effectively working as a single control qubit \cite{Mottonen2004,Mottonen2005,Babbush2018}. The control qubits then constitutes a switch which can either be in an \enquote{open} state, which, in the case of four qubits, i.e., two control qubits, is $|0\rangle_C \equiv |00\rangle_C$, or a \enquote{closed} state, which, in this case is the Bell states $|1^{\pm}\rangle_C = (|10\rangle_C \pm |01\rangle_C)/\sqrt{2}$, depending on the choice of $\Delta_\pm$. Note that the subscript $C$ denotes the $(N-2)$-qubit state of the control qubits, while we use $T$ for the target, i.e., first and last, qubits. In the computational basis of the target qubits, $\{|00\rangle_T, |01\rangle_T,|10\rangle_T,|11\rangle_T\}$, the open gate can be expressed as 
\begin{equation}\label{eq:Uopen}
	U_\text{open} = \begin{pmatrix}
	1 & 0 & 0 & 0 \\
	0 & 0 & \mp 1 & 0 \\
	0 & \mp 1 & 0 & 0 \\
	0 & 0 & 0 & i
	\end{pmatrix},
\end{equation}
where the choice of $\Delta_\pm$ dictates the phase on the swap. The closed state of the gate is simply the identity $U_\text{closed} = \unitm_4$. The open gate will entangle the input and output qubits. This can be quantified using the entanglement power \cite{Zanardi2000}, which in our case is 1/9. 

\begin{figure*}
	\centering
	\includegraphics[width=\textwidth]{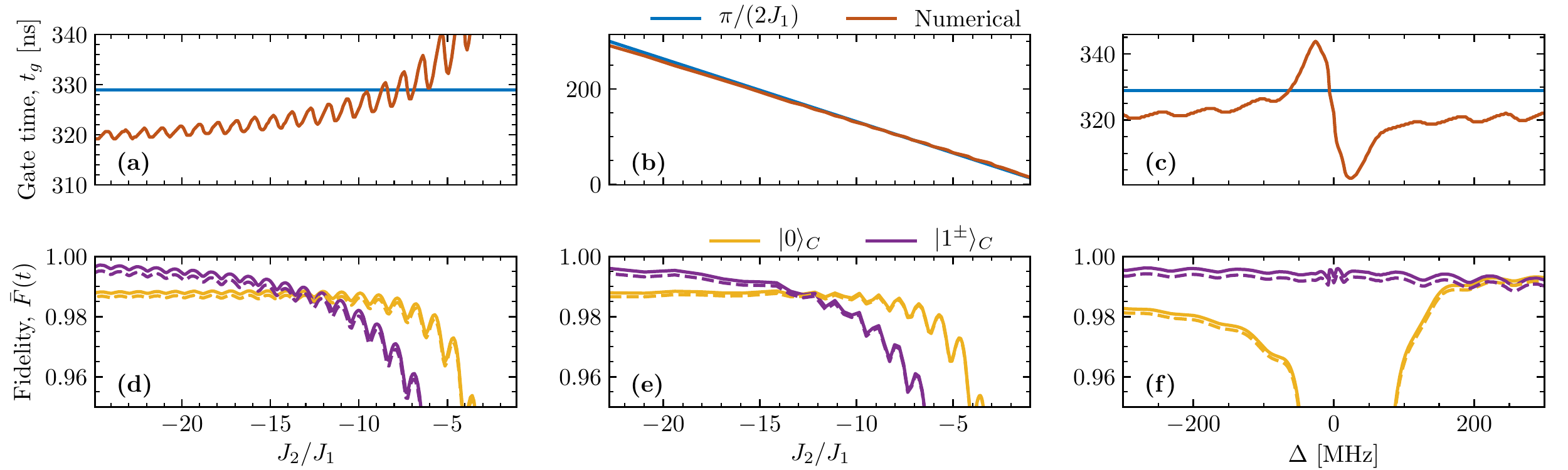}
	\caption{[\textbf{(a)-(c)}] Gate time as a function of the model parameters $J_2$, $1/J_1$, and $\Delta$. The blue lines indicates the analytical result of \cref{eq:gateTime} and the red lines indicates the point of maximum average fidelity. [\textbf{(d)-(e)}] Average fidelities at the numerical gate time as a function of the model parameters $J_2$, $1/J_1$, and $\Delta$, both with (dashed lines) and without (solid lines) decoherence noise. The yellow lines indicates the fidelity when the gate is in the open configuration, while the purple line indicates that it is in the closed configuration.}
	\label{fig:scan}
\end{figure*}

In order to quantify the effectiveness of the gate, we use numerical simulations with realistic gate parameters for state-of-the-art superconducting circuits. In \cref{fig:Qutrit}, we show simulations of the gate in a circuit with specific superconducting circuit parameters. We include decoherence noise occurring in superconducting circuits by considering the Lindblad master equation 
\begin{equation}\label{eq:linblad}
\dot{\rho} = -i [H,\rho] + \gamma\sum_{j}  \left[ A_l \rho A_l^\dagger - \frac{1}{2} \lbrace \rho, A_l^\dagger A_l\rbrace \right],
\end{equation}
where $\rho$ is the density matrix, $H$ is the Hamiltonian in \cref{eq:HI}, the curly brackets indicates the anticommutator, and the sum is taken over the eight collapse operators $A_l$: $\sigma_j^z$ inducing dephasing, and $\sigma_j^-$ inducing photon loss, with $j$ running over all qubits.  We take the decoherence rate, $\gamma$, to be identical on all qubits with a state-of-the-art rate of $\gamma = \SI{0.01}{\MHz}$, giving the qubits a lifetime of $\SI{100}{\micro\second}$ \cite{Wang2018}.

In order to measure the quality of the gate, we consider the average fidelity \cite{Nielsen2002}
\begin{equation}
	\bar{F}(t) = \frac{1}{5} + \frac{1}{80} \sum_{j=1}^{16} \text{Tr} \left(U_\text{target} U_j^\dagger U_\text{target}^\dagger \mathcal{E}_t(U_j)\right),
\end{equation}
which evaluates how well a quantum map, $\mathcal{E}_t$, approximates the target gate, $U_\text{target}$, over a uniform distribution of input quantum states, $U_j$. The target operator is either $U_\text{open}$ or $U_\text{closed}$ depending of the state of the control qubits, which is encoded in the initial density matrix $\rho(0)$. By solving the Lindblad master equation, \cref{eq:linblad} we obtain the density matrix at a later time, $\rho(t)$. This is done using the \textsc{Python} toolbox \textsc{ QuTiP} \cite{QuTip2013}. Having obtained the full density matrix we can then trace out the control degrees of freedom, yielding the desired quantum map $\mathcal{E}_t(\rho(0)) = \text{Tr}_T(\rho(t))$. We chose the basis, $U_j$, of the average fidelity as all two-qubit Pauli operators on the form $(\sigma_1^x)^k(\sigma_1^z)^l(\sigma_N^x)^m(\sigma_N^z)^n$ for all combinations of $k,l,m,n \in \{0,1\}$.

Thus given a set of model parameters, the average fidelity can be calculated as a function of time for both set of gate configurations. In the case of the open gate, i.e., configuration $|0\rangle_C$, the average fidelity rises from some initial value to a maximum (unity for the perfect gate) at the gate time, which we denote $t_g$. Analytically, we expect this to be (see Supplementary Material \cite{supplMat} for a derivation of this)
\begin{equation}\label{eq:gateTime}
	t_g = \frac{\pi}{|2J_1|};
\end{equation}
however, for the simulations we find the best gate time numerically. In the case of the closed gate, i.e., the configuration $|1\rangle_C$, the average fidelity is initially unity and deviates only from this value due to leakage to the control qubits or as a result of decoherence noise.

In order to investigate the sensitivity of the parameter space, we vary the parameters $J_1$, $J_2$, and $\Delta$ and show the gate time and average fidelities at the gate time in \cref{fig:scan}. The simulation is done both with and without noise. In Figs. \ref{fig:scan}(a) and (d), we vary the coupling of the control qubits, $J_2 \equiv J^x_2 = J^z_2$, in the configuration $|1^+\rangle_C$, while keeping the remaining coupling constants at $J_1 = \SI{30}{\MHz}$. Setting $J_2^x = J_2^z$ is merely done for the simplicity of the numerical investigation and is not a requirement, as we will exploit later. From this simulation we observe that the numerical gate time is about 5\% faster than the analytical, and for large $J_2/J_1$ we observe almost unity average fidelity for the closed configuration of the gate, and between 0.98 and 0.99 for the open configuration. In Figs. \ref{fig:scan}(b) and (e), we vary the coupling between the target qubits and the control qubits, i.e. $J_1$, while keeping the coupling between the control qubits constant at $J_2 = \SI{750}{\MHz}$, in the configuration $|1^+\rangle_C$. Again, we observe a slightly shorter numerical gate time, and fidelities of close to unity and just between 0.98 and 0.99 for the closed and open configuration respectively. In Figs. \ref{fig:scan}(c) and (f), we vary $J_2^x$ and keeping $J_2^z = \SI{600}{\MHz}$ in the case of the gate being in the configuration $|1^-\rangle_C$ in order to effectively vary $\Delta$ around zero. We observe that the gate completely fails around zero, as it should, but we also conclude that we achieve a larger average fidelity (just above 0.99) for a positive detuning, i.e., $J_2^z > J_2^x$, rather than a negative detuning. However, for the case of $|1^+\rangle_C$, we find that the average fidelity is slightly larger when $J_2^z < J_2^x$. From the simulations, we also find that a different sign on the couplings $J_1$ and $J_2$ yields a slightly larger average fidelity.

The above mentioned simulations beg the question of why the average gate fidelities do not approach unity, even when the requirements mentioned in \cref{sec:H} are fulfilled. The answer to this question is found together with the answer as to why the numerical gate time is shorter than the analytical gate time in \cref{eq:gateTime}. It all comes down to the fact that even though the state $|1\rangle |0\rangle_C |1\rangle$ is indeed an eigenstate of the Hamiltonian, it is also degenerate with the states $|1\rangle |1^\mp\rangle_C |0\rangle$ and $|0\rangle |1^\mp\rangle_C |1\rangle$ depending on the choice of $\Delta_\pm$ (note that these states are not the same as the configurations of the closed gate). This means that the system will oscillate between these three states, in a manner similar to how the open gate oscillates between states with a single excitation. However, the time scale for the oscillation of the double excitation is less than for the single excitation, with an oscillation time of about 90\% of the analytical gate time. This means that some time between $0.9t_g$ and $t_g$ we will observe maximum average fidelity, less unity, depending on the configuration of the system. 

This does, however, not mean that it is impossible to achieve perfect transfer for some states, in a well configured system. Namely as long as not \emph{both} the input and output qubit are in a superposition state, the state is transfered perfectly, when disregarding decoherence noise.

Note that the resonance of the eigenstates mentioned above is the same resonance that makes the gate work to begin with, in that case it is the states $|1\rangle |0\rangle_C |0\rangle$, $|0\rangle |0\rangle_C |1\rangle$, and $|0\rangle |1^\mp\rangle_C |0\rangle$ that are in resonance.

\section{Possible physical realization}\label{sec:realization}

We wish to implement the Hamiltonian in \cref{eq:HI}, and thus the swapping gate, using superconducting circuits. As in the previous section, we will focus on implementing the case of $N=4$, but the idea is easily expanded to larger $N$.

\subsection{Superconducting circuit}\label{sec:SCC}

\begin{figure}
	\centering
	\includegraphics[width=\columnwidth]{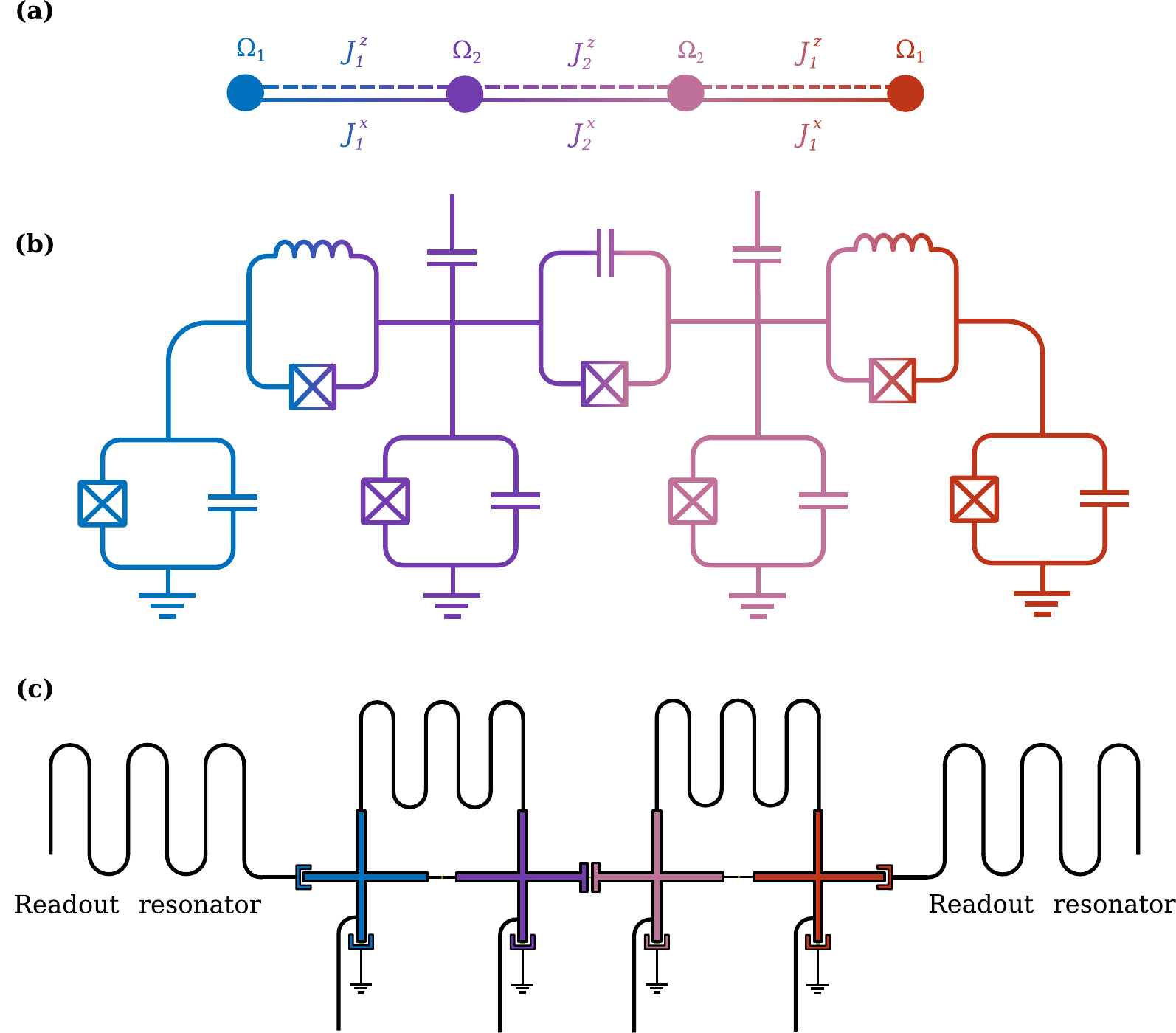}
	\caption{\textbf{(a)} The desired spin model as seen in \cref{eq:HI}. \textbf{(b)} The lumped circuit model for the super conducting circuit used to implement the above system. The crossed boxes represent Josephson junctions, the parallel lines are capacitors, and the curled wires are inductors. \textbf{(c)} Possible circuit design, consisting of four X-mon-style superconducting islands, all grounded and with control lines from below. The small green patches indicate Josephson junctions, while the bent wires represent inductors. The first and last qubits are connected to a readout resonator as well. The colors indicates which parts of the three models correspond to each other.}
	\label{fig:SCCmodel}
\end{figure}

The circuit used to implement the system can be seen in \cref{fig:SCCmodel}(b). The circuit consists of four Transmon-type qubits \cite{Transmon2007}, which are all grounded and connected in series through Josephson junctions, with as small of a parasitic capacitance as possible. In parallel with the connecting Josephson junctions is either a capacitor or an inductor, alternating between these two. Additional qubits are added to the chain by connecting them through a Josephson junction and either a capacitor or an inductor. It is important that two connecting capacitances are not next to each other, since this will induce cross talk between the nodes. When there is only a capacitor between every other pair of nodes the capacitance matrix becomes block diagonal, which means that its inverse will be block diagonal as well. However, had there been capacitors between all nodes the capacitance matrix would have been tridiagonal, and its inverse would not necessarily be tridiagonal, which possibly yields cross talk, i.e., couplings other than nearest neighbor coupling. In reality, there will always be a parasitic capacitance between two nodes connected through a Josephson junction; however, not including these are equivalent to assuming $C_i \gg C_{i,i+1}$, where $C_i$ is the shunting capacitance of the $i$th qubit and $C_{i,i+1}$ is the parasitic capacitance between the $i$th and $(i+1)$th node. See \cref{app:SSH} for a discussion on the emergence of cross talk due to capacitive couplings.

Instead of transmon qubits one could, in principle, have used other types of superconducting qubits such as the C-shunted qubit (or floxmons) \cite{Orlando1999,Mooij1999,Wal2000,You2007,Yan2016} or fluxonium \cite{Manucharyan2009}. 

For each node in the circuit, we have a related flux degree of freedom, which we denote $\phi_i$ \cite{Devoret2017}. Interactions between the qubits are induced by capacitors and inductors, which induce $XX$ couplings, and Josephson junctions which induce both $XX$ and $ZZ$ couplings. A detailed calculation going from the circuit design to the Hamiltonian in \cref{eq:HI} can be found in the Supplementary Material \cite{supplMat}. Since Josephson junctions are nonlinear inductors and thus induce both $XX$ and $ZZ$ couplings between the qubits, it might seem redundant to include inductors or capacitors in parallel with these. However, these are included such that it is possible to tune the $XX$ coupling without affecting the $ZZ$ coupling significantly.

In order to operate the gate successively, i.e., opening and closing the gate in an uninterrupted sequence, we need a scheme for preparing the state of the gate. We would like to be able to address the control qubits exclusively, i.e., opening and closing the gate independently of the target qubits. This is possible when the target qubits are detuned from the control qubits, i.e., $\Delta$ is sufficiently large, compared to the couplings between the control qubits and the target qubits. A large detuning can, in experiments, be obtained by tuning the external fluxes.

\begin{figure}
	\centering
	\includegraphics[width=0.8\columnwidth]{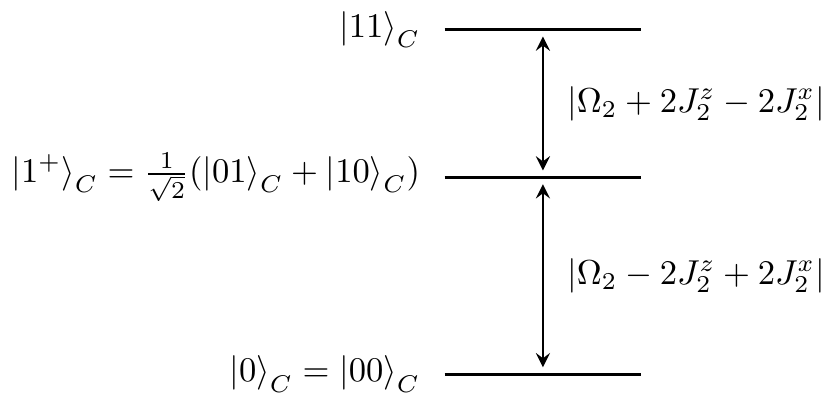}
	\caption{Sketch of the state of the control qubits. The states of the target qubits have not been added since $J_1 \ll J_2$, and the fact that we wish to operate the gate on a timescale where the target qubits are stationary i.e. irrelevant for the difference in energy of the gate's energy levels. One can further detune the gate from the target qubits by using the flux lines depicted in \cref{fig:SCCmodel}, thus effectively eliminating the coupling between the gate and the target qubits.}
	\label{fig:tripletGateState}
\end{figure}

We can achieve control of the gate by driving the middle nodes. The driving is performed by adding an external field to the nodes through capacitors. The control lines are depicted in \cref{fig:SCCmodel}(c) as the wires left of the ground. This introduces an extra driving term to the Hamiltonian, which in the interaction picture takes the form
\begin{equation}
H_d(t) = \frac{A}{2}I(t) \left[  (\sigma_2^y + \sigma_3^y)\cos \delta t  - (\sigma_2^x + \sigma_3^x)\sin \delta t  \right],
\end{equation} 
for an in-phase driving, where we have defined $\delta = \omega-\Omega_1$, $\omega$ is the driving frequency, and $I(t)$ is the envelope of the driving pulse. 
Like the rest of the Hamiltonian this term preserves the total spin of the two gate qubits, and hence it does not mix the singlet and triplet states. We can therefore ignore the singlet state $\left|1^- \right\rangle_C$, when starting from any of the triplet states shown in \cref{fig:tripletGateState}. 

Rabi oscillations between the closed and open states are then generated by the driving, provided the driving frequency matches the energy difference $\omega = |\Omega - 2J_{2}^z + 2J_{2}^x|$ and $A\ll J_{2}^z$.
A $\pi$ pulse would then shift between the $\left| 0 \right\rangle_C$ and $\left| 1^+ \right\rangle_C$ states in a few microseconds depending on the size of $A$.
The energy difference between the open or closed states and the last state $\left| 11 \right\rangle_C$ are far enough from $\omega$ such that we do not populate this state by accident. Thus, using this scheme, we can drive between an open and closed gate using merely an external microwave drive. For a detailed calculation of the driving force see the Supplementary Material \cite{supplMat}.

If we were to drive the system for an intermediate time between zero and one $\pi$ pulse, we would obtain a superposition of the open and closed gate. Suppose that we drive the Rabi oscillation for half a $\pi$ pulse, $t=\pi/2A$: In this case we would get the superposition
\begin{equation}\label{eq:SuperGate}
|1^+\rangle_C \rightarrow \frac{1}{\sqrt{2}}\left(|1^+\rangle_C + i|0\rangle_C \right).
\end{equation}
In this case the gate would permit a superposition of the system being transferred and not. In the same way that the transferred state accumulates a phase during the transfer, so does the superposition gate. The phase obtained by the superposition gate is simply the energy difference between the open and closed state. Thus, we must include a phase factor of
\begin{equation}
e^{-i(\Omega_2 - 2J_{2}^z + 2J_{2}^x)t}
\end{equation}
on the gate when evaluating the state.

A lumped circuit diagram is not enough for a possible realistic implementation and we therefore propose an experimental realistic chip design, which can be seen in \cref{fig:SCCmodel}(c). The chip consists of four X-mon-like superconducting islands \cite{Barends2013}, each connected to the ground through a Transmon qubit, and each connected to a control line. All qubits are connected to its neighbor through a Josephson junction, and the two middle are close to one another in order to create a capacitive coupling, while the outer islands are farther from the middle islands in order to minimize the capacitive coupling, while being connected through an inductor each. The outer islands corresponds to the target qubits and are therefore connected to an LC resonator each, in order to be able to perform measurements on them. The two middle islands corresponds to the control qubits.
 
\subsection{Leakage and infidelities}\label{sec:leakage}

\begin{figure}
	\centering
	\includegraphics[width=\columnwidth]{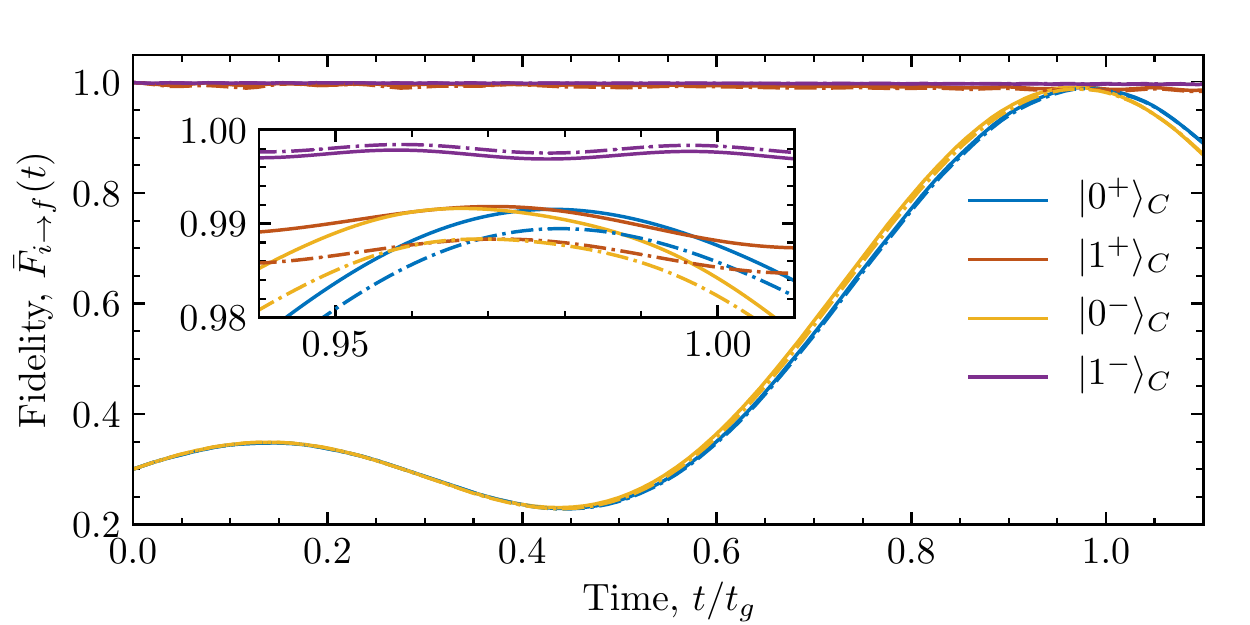}
	\caption{Average fidelity of the system with realistic parameters as a function of time. The simulations are done both for control qubits (solid lines) and qutrits (dashed dotted lines). The average fidelity is plotted for all possible configurations of the control qubits and qutrits. The $|0^\pm \rangle_C$ indicates the choice of detuning $\Delta_\pm$ and thus which of the open gates in \cref{eq:Uopen} the simulation is done for. The inset shows a zoom of the peak of the average fidelity in the open configuration $|0\rangle_C$, i.e., around $t \sim t_g$.}
	\label{fig:Qutrit}
\end{figure}

Because that we require a large coupling coefficient between the two control qubits, $J_2$, the superconducting circuit is vulnerable to leakage to higher excited states than the two lowest states. In order to investigate the amount of leakage in our system we simulate it with the control qubits being qutrits, i.e., including the three lowest states of the system. The simulation is done for realistic parameters which can be found in Table S1 of the Supplementary Material \cite{supplMat}. The average fidelity is shown in \cref{fig:Qutrit}, together with the results for the system with purely qubits. 

From \cref{fig:Qutrit}, we observe that the average fidelity is not affected significantly in the large picture; however, if we zoom in on the peak of the average fidelity in the open state, we observe that the average fidelity has indeed decreased in both cases of the open gate and the case of the $|1^+\rangle$-closed gate, while the fidelity of the $|1^-\rangle$-closed case has increased. In general, whether the inclusion of the second excited state in the gate will increase or decrease the fidelity is dependent on the parameters of the gate. However, as long as the anharmonicity remains large ($\mathcal{A}^r \gtrsim 0.1 \%$ \cite{Transmon2007}) the effect will be minimal, not changing the overall picture.

\begin{figure}
	\centering
	\includegraphics[width=\columnwidth]{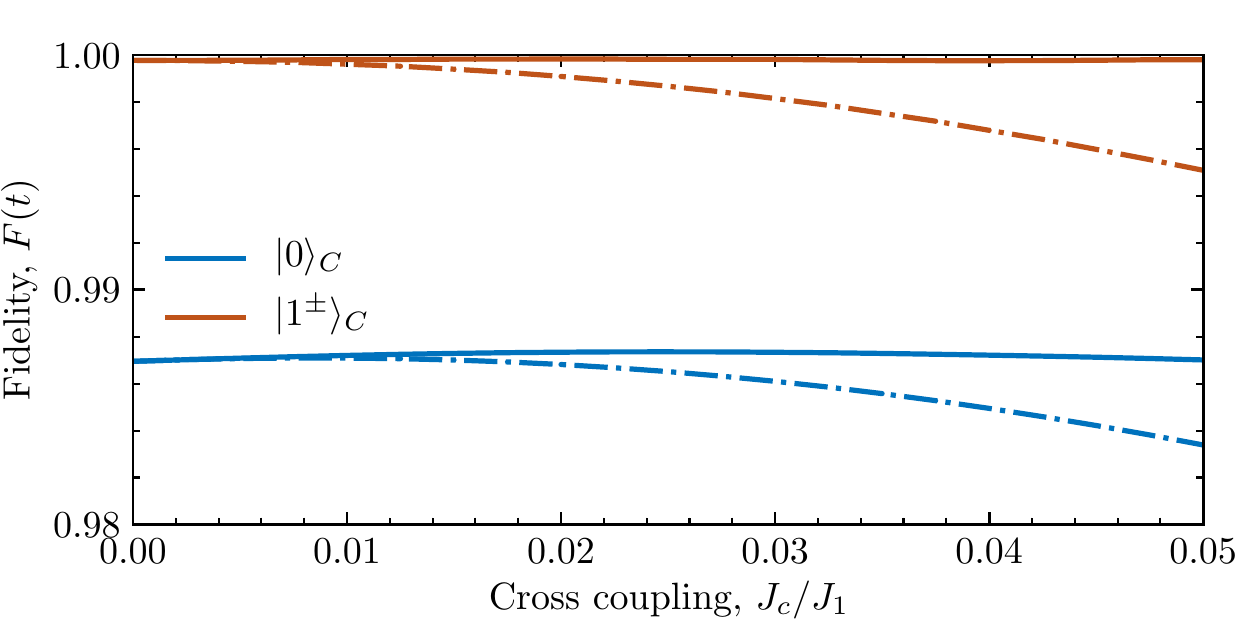}
	\caption{Average fidelities of the control qubits in both states for increasing strength of coupling beyond nearest neighbor coupling, $J_c$. The solid lines is with next-to-nearest couplings included, and the dash-dotted lines are with up to next-to-next-to-nearest neighbor coupling.}
	\label{fig:crossTalk}
\end{figure}

Besides leakage to higher excited states capacitive couplings beyond nearest neighbor qubits possess the biggest threat to gate fidelity. We therefore consider the effect of cross talk on the average gate fidelities as seen in \cref{fig:crossTalk}. From the simulation, we see that next-to-nearest couplings have no effect on the gate when it is in its closed configurations $|1^\pm\rangle_C $, and have only little effect when it is in its open configuration $|0\rangle_C$. In fact, the average fidelity increases a tiny amount until the next-to-nearest neighbor coupling is $\sim3\%$ of the  nearest neighbor coupling between the target qubits and the control qubits. This is consistent with the result of Ref. \cite{Loft2018b}. Next-to-next-to-nearest couplings, on the other hand, have a much more significant influence on the system when the coupling strength is above 2\% of the target-control coupling, and we conclude that the gate fidelity decreases as the square of the next-to-next-to-nearest coupling strength. This is expected since the next-to-next-to-nearest coupling is a direct coupling of the input and output qubits.

\section{Extensions and outlook}\label{sec:circuitOutlook}

Besides being used for the above mentioned swapping gate, the superconducting circuit is interesting in many other settings because it implements the most fundamental of all spin model structures: the linear spin chain. The linear chain is the obvious choice for a \textquote{quantum wire} in an implementation of quantum information processing, especially if configured for perfect state transfer over a fixed period of time \cite{Christandl2004}. The superconducting circuit in \cref{fig:SCCmodel} is a possible candidate for this application because of its straight forward scaling.

Consider the case where we want a model without any $ZZ$ couplings. One way to achieve this is simply to fine-tune the system and thus suppress the $ZZ$ couplings. However, there is an easier way: All the contributions to the $ZZ$ coupling stem from the Josephson junctions, and thus removing those will create a purely $XX$-coupled spin chain. One could even remove the capacitors and just couple the qubits through a series of inductors similar to the chain in Ref. \cite{Neill2018} or the one-dimensional tight-binding lattice for photons mentioned in Ref. \cite{Ningyuan2015}, but with superconducting qubits instead of LC resonators in order to create a spin model and not a boson model. This circuit could also be used to investigate the Su-Schrieffer-Heeger (SSH) model \cite{Su1979,Heeger1988} defined on the dimerized one-dimensional lattice with two sites per unit cell, both in the strong and weak coupling limit in relation the the Zak phase as considered by Ref. \cite{Goren2018}. It should be mentioned that we were not able to reproduce the lattice in Ref. \cite{Goren2018} using their suggested circuit because of the previously mentioned fact that the inverse of the capacitance matrix, with couplings entirely with capacitors, is not tridiagonal, which induces non-negligible cross talk, especially in the strong coupling limit. The occurrence of this problem is illuminated in \cref{app:SSH}. Our circuit does not introduce this cross talk, and is therefore more suitable for the investigation of the SSH model.

The superconducting circuit presented here can also easily be molded into a box model where each qubit corresponds to a corner of the box and each edge a coupling. This is done simply by connecting the first and last superconducting island of the circuit, i.e., the blue and red islands in \cref{fig:SCCmodel}, with a Josephson junction and/or a capacitor depending on which kind of coupling one wishes to implement. In a realistic implementation, it might be necessary to place the X-mon superconducting island in a square pattern instead of a linear pattern, as seen in \cref{fig:SCCmodel}(c). Such a system could be used to engineering quantum spin liquids and many-body Majorana states \cite{Yang2018}.

Lastly, we mention that even though we have tried to avoid cross talk up until now, it is possible to modify the circuit into having all-to-all $XX$ couplings by connecting all superconducting islands with capacitors. The most effective implementation of this would be in the box shape, since this avoids the use of 3D integration \cite{Rosenberg2017,Lu2018}. For all-to-all $ZZ$ couplings, one would have to use 3D integration as all superconducting islands must be connected directly via Josephson junctions.

\section{Conclusion}\label{sec:conclusion}

We have investigated a linear chain of qubits and found that under the right configuration these can operate as a two-qubit swapping gate with different phases depending on the initial conditions. In particular we have focused on the case of four qubits and shown that it can create a swapping gate with a fidelity around 0.99, even when including realistic decoherence noise and leakage to higher excited states.

Furthermore, we have proposed a superconducting circuit which realizes the four-qubit spin chain. Both a lumped circuit model and a possible chip design using X-mon-style qubits have been presented and we have discussed how to operate this circuit between the different configurations of a given gate. Finally, we have discussed how the superconducting circuit can be modified in a simple manner to realize other models with Hamiltonians that have attracted considerable interest in recent times. This shows that the basic model and layout we propose may have extended utility in both quantum processing and quantum simulation research directions.

\begin{acknowledgments}
The authors would like to thank T. Bækkegaard, L. B. Kristensen, and N. J. S. Loft for discussion on different aspects of the work.
This work is supported by the Danish Council for Independent Research and the Carlsberg Foundation.
\end{acknowledgments}

\appendix
\section{Full capacitive couplings}\label{app:SSH}

In this appendix, we consider different cases of capacitive couplings between qubits. We consider both cases which yield couplings beyond nearest neighbor couplings and cases which yields only nearest neighbor couplings, as is desired in our case.

\begin{figure}
	\centering
	\includegraphics[width=\columnwidth]{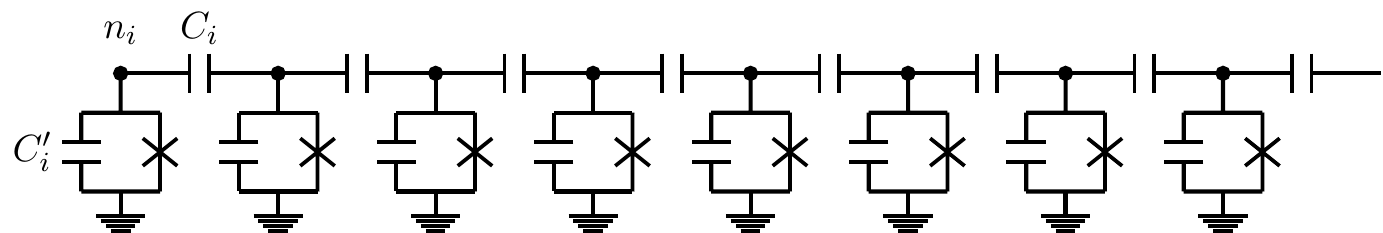}
	\caption{The circuit consists of $N$ qubits connected through capacitors of size $C_i$. Each transmon consists of a capacitor of size $C'_i$ and a Josephson junction (usually a SQUID) of size $E_i$. The nodes of the circuit are denoted $n_i$.}
	\label{fig:SSHcircuit}
\end{figure}

Consider the circuit in \cref{fig:SSHcircuit}, which is $N$ qubits in series coupled with capacitors. This circuit yields a Lagrangian of
\begin{equation}
\begin{aligned}
L =& \sum_{n=1}^{N} \left( \frac{1}{2}C_i\dot{\phi}_i^2  + E_i\cos \phi_i \right),
\end{aligned}
\end{equation}
which yields a Hamiltonian of 
\begin{equation}
\begin{aligned}
H =& 4 \vec{p}^T K^{-1} \vec{p} -\sum_{n=1}^{N}E_i\cos \phi_i.
\end{aligned}
\end{equation}
Now consider that we want identical couplings between all of the qubits: We therefore set $C_i =C'_i = C$, which yields the following inverse capacitance matrix here for the simple case of $N=4$:
\begin{equation}\label{eq:invKN4}
	K^{-1} = \frac{1}{C} \begin{pmatrix}
	1 & 0 & -1 & 1 \\
	0 & 0 & 1 & -1 \\
	-1 & 1 & 0 & 0 \\
	1 & -1 & 0 & 1 
	\end{pmatrix}.
\end{equation}
From this, we see that the desired nearest neighbor coupling simply disappears between some of the qubits, while couplings beyond nearest neighbor are significant. Increasing the number of qubits does not fix this. In fact, in all cases where $N+1$ is dividable with 3, the matrix is even singular. One could try to fix this simply by increasing size of the shunting capacitor $C'_i$ of the qubit compared to the coupling capacitor $C_i$. Thus, for $C_i' = 10 C_i = 10 C$, we get
\begin{equation}
K^{-1} \simeq \frac{1}{C} \begin{pmatrix}
1 & -0.1 & 0.01 & -0.001 \\
-0.1 & 1 & -0.1 & 0.01 \\
0.01 & -0.1 & 1 & -0.1 \\
-0.001 & 0.01 & -0.1 & 1 
\end{pmatrix},
\end{equation}
which does indeed seem to fix the problem. Consider now the case where we want to make a SSH chain, such as in Ref. \cite{Goren2018}. In this case, we alternate the coupling between the qubits such that in the case of $N=6$
\begin{equation}
K = \begin{pmatrix}
C'_1 & C & 0 & 0 & 0 & 0 \\
C & C'_2 & 2C & 0 & 0 & 0 \\
0 & 2C & C'_3 & C & 0 & 0 \\
0 & 0 & C & C'_4 & 2C & 0 \\
0 & 0 & 0 & 2C & C'_5 & C \\
0 & 0 & 0 & 0 & C & C'_6 \\
\end{pmatrix}.
\end{equation}
For $C'_i = C$, we obtain a result similar to \cref{eq:invKN4} in the first example where some of the nearest neighbor couplings disappear, and couplings beyond this become large. This can be fixed using the approach mentioned above with $C \ll C_i'$.
The proposal of Ref. \cite{Goren2018} suggests to take $C \gg C_{i}^{'}$ in order to enter the strong-coupling regime and realize a linear SSH chain. Given the above analysis, we cannot see how this is feasible without some other modifications to the circuit.  

In order to avoid couplings beyond nearest neighbor couplings, the superconducting qubits should instead be connected with inductors or connected with alternating inductors and capacitors in order to avoid cross talk. However, if one desires a SSH model, it is still not enough to alternate between the sizes of the inductors or capacitors if the chain is of finite length, due to end-point irregularities.

If the circuit is constructed with only capacitors on every other coupling between the qubits the capacitance matrix becomes block diagonal. Consider a block-diagonal matrix consisting of invertible matrices $\mathbf{K}_i$
\begin{equation}
	K = \begin{pmatrix}
	\mathbf{K}_1 & 0 & 0 &\cdots & 0 \\
	0 & \mathbf{K}_2 & 0 & \cdots & 0 \\
	0 & 0 & \mathbf{K}_3 & \cdots & 0 \\
	\vdots & \vdots & \vdots & \ddots &  0 \\
	0 & 0 & 0 & 0 & \mathbf{K}_N \\
	\end{pmatrix},
\end{equation}
which yields an inverse capacitance matrix of
\begin{equation}
K^{-1} = \begin{pmatrix}
\mathbf{K}_1^{-1} & 0 & 0 &\cdots & 0 \\
0 & \mathbf{K}_2^{-1} & 0 & \cdots & 0 \\
0 & 0 & \mathbf{K}_3^{-1} & \cdots & 0 \\
\vdots & \vdots & \vdots & \ddots &  0 \\
0 & 0 & 0 & 0 & \mathbf{K}_N^{-1} \\
\end{pmatrix},
\end{equation}
which is block diagonal as well. If the matrices $\mathbf{K}_i$ are $2\times 2$ the capacitance matrix yields only nearest neighbor couplings. The lack of couplings between the blocks of the matrix can be fixed by adding inductors between the blocks of qubits.

In the specific case of $N=4$ qubits discussed in the main text the capacitance matrix is given as
\begin{equation}
K = \begin{pmatrix}
C_1 & 0 & 0 & 0 \\
0 & C_2 + C_{2,3} & -C_{2,3} & 0 \\
0 & -C_{2,3} & C_2 + C_{2,3} & 0 \\
0 & 0 & 0 & C_1 
\end{pmatrix},
\end{equation}
which yields an inverse capacitance matrix of
\begin{align}
K^{-1} =\begin{pmatrix}
1/C_1 & 0 & 0 & 0 \\
0 & (C_2+C_{23})/C_0 & -C_{2,3}/C_0 & 0 \\
0 & -C_{2,3}/C_0 & (C_2+C_{2,3})/C_0 & 0 \\
0 & 0 & 0 & 1/C_1 
\end{pmatrix},
\end{align}
where $C_0 = C_2^2 + 2C_{2,3}C_2$. This yields only couplings between the middle two qubits.

%\bibliography{references}


\end{document}

\title{Supplementary Material for:\\A controllable two-qubit swapping gate using superconducting circuits}

\author{S. E. Rasmussen}
\email{stig.elkjaer.rasmussen@post.au.dk}
\affiliation{\affA}
\author{K. S. Christensen}
\affiliation{\affA}
\author{N. T. Zinner}
\email{zinner@phys.au.dk}
\affiliation{\affA}
\affiliation{\affB}

\maketitle

\section{The requirements for the gate}\label{app:derivReq}

This section roughly follows the work of Ref. \cite{Marchukov2016}, but uses a different approach in some instances and includes a derivation of the requirement for perfect state transfer of superposition states.
Consider a chain of $N$ spin-1/2 particles described by a Heisenberg $XXZ$ spin model as in Eq.~(3) of the main text, with the same spatial symmetry requirements, i.e. $J_i^{x,z} = J_{N-i}^{x,z}$ and $\Delta_i = \Delta_{N-1-i}$.

\subsection{Deriving the relevant eigenstates}\label{app:QSTN4}

Since the Hamilton in Eq.~(3) preserves excitation we can consider the problem in each subspace, $\mathcal{B}_k$ of total excitation, $k = 0,\dots,4$. A closed state of a gate allows no dynamics, thus we require the state to be stationary. This is achieved by the eigenstate of the Hamiltonian. For the sake of completeness consider first the subspaces $\mathcal{B}_{0,4}$ these consist of only one state, $\ket{0000}$ or $\ket{1111}$, these are obviously the eigenstate of the system and stationary (since the Hamiltonian is preserves excitation we already knew this). The eigenenergies of these states are $E_{0,4} = \mp \Delta + J_2^z + 2J_1^z$, minus for $E_0$ and plus for $E_4$.

Consider now the subspaces $\mathcal{B}_{1}$ consisting of the states $\{ \ket{1000},\allowbreak \ket{0100},\allowbreak \ket{0010},\allowbreak \ket{0001} \}$ (the subspace $\mathcal{B}_{3}$ works in an identical way just with the excitation of every state exchanged, i.e. exchanging $0 \leftrightarrow 1$ in the states). In this basis the Hamiltonian matrix reads
\begin{equation}\label{eq:HmatN4}
H_{1} = \begin{pmatrix}
-\Delta + J_2^z & 2J_1^x & 0 & 0 \\
2J_1^x & - J_2^z & 2J_2^x & 0 \\
0 & 2J_2^x & - J_2^z & 2J_1^x \\
0 & 0 & 2J_1^x & -\Delta + J_2^z
\end{pmatrix}.
\end{equation}
Now consider the last and largest subspace $\mathcal{B}_2$ consisting of the six states $\{ \ket{0011},\allowbreak \ket{0101},\allowbreak \ket{0110},\allowbreak \ket{1001},\allowbreak
\ket{1010},\allowbreak
\ket{1100} \}$. In this basis the Hamiltonian matrix reads
\begin{align}\label{eq:H0}
&H_2 = \left(\begin{smallmatrix}
2J_1^z -J_2^z & 2J_2^x & 0 & 0 & 0 & 0 \\
2J_2^x & -2J_1^z - J_2^z & 2J_1^x &  2J_1^x & 0 & 0  \\
0 & 2J_1^x & - 2J_1^z + J_2^z - \Delta & 0 & 2J_1^x & 0 \\
0 & 2J_1^x & 0 & - 2J_1^z + J_2^z + \Delta & 2J_1^x & 0 \\
0 & 0 & 2J_1^x & 2J_1^x & -2J_1^z - J_2^z & 2J_2^x \\
0 & 0 & 0 & 0 & 2J_2^x & 2J_1^z -J_2^z
\end{smallmatrix}\right).
\end{align}
Now we wish for the closed state to be a superposition of the the two middle qubits when they have one excitation combined, thus we require the following the two states to be eigenstates of the Hamiltonian
\begin{subequations}
	\begin{align}
	\ket{\psi_1} &= \cos\theta \ket{0100} + \sin \theta  \ket{0010}, \\
	\ket{\psi_2} &= \cos\theta \ket{1100} + \sin \theta  \ket{1010}.
	\end{align}
\end{subequations}
Thus applying the Hamiltonian to the states we obtain
\begin{align*}
H_{1}\ket{\psi_1} &= \begin{pmatrix}
2J_1^x \cos\theta \\
-J_2^z \cos\theta + 2J_2^x \sin \theta \\
2J_2^x \cos\theta - J_2^z \sin \theta \\
2J_1^x \sin\theta
\end{pmatrix} = b_1 \begin{pmatrix}
0 \\
\cos\theta \\
\sin\theta \\
0
\end{pmatrix}, \\
H_{2}\ket{\psi_2} &= \begin{pmatrix}
(2J_1^z -J_2^z) \cos\theta + 2J_2^x\sin\theta \\
2J_2^x \cos\theta - (2J_1^z + J_2^z) \sin \theta \\
2J_1^x \sin \theta \\
2J_1^x \sin\theta \\
0 \\ 
0
\end{pmatrix} = b_2 \begin{pmatrix}
\cos\theta \\
\sin\theta \\
0 \\
0 \\
0 \\
0 
\end{pmatrix},
\end{align*}
where the last equality is the eigenstate requirement. For these equations to be satisfied it is evident that $J_1$ must vanish for both the $x$ and $z$ couplings. However, setting these coefficients to zero  would decouple the middle qubits from the end qubits, and thus removing any dynamics of the system. We therefore settle for requiring $J_1 \ll J_2$.
From the remaining equations we see that $\theta = \pm \pi/4$ (not surprising considering the symmetry of the problem). This yields $b_1 = b_2 = 2J_2^x - J_2^z$. 

Having found two eigenstate for $H_{1}$ ($\theta = \pm \pi/4$), we make a unitary transformation to the basis where these are eigenstates using the transformation matrix
\begin{equation}
\mathcal{V} = \begin{pmatrix}
1 & 0 & 0 & 0 \\
0 & \frac{1}{\sqrt{2}} & \frac{1}{\sqrt{2}} & 0 \\
0 & \frac{1}{\sqrt{2}} & -\frac{1}{\sqrt{2}} & 0 \\
0 & 0 & 0 & 1
\end{pmatrix},
\end{equation}
which yields
\begin{equation}
\tilde{H}_{1} = \mathcal{V}^{-1} H_{1}\mathcal{V} = \left(\begin{smallmatrix}
-\Delta + J_2^z & \sqrt{2}J_1^x & \sqrt{2}J_1^x & 0 \\
\sqrt{2}J_1^x & 2J_2^x - J_2^z & 0 & \sqrt{2}J_1^x \\
\sqrt{2}J_1^x & 0 & -2J_2^x - J_2^z & -\sqrt{2}J_1^x \\
0 & \sqrt{2}J_1^x & -\sqrt{2}J_1^x & -\Delta + J_2^z
\end{smallmatrix}\right),
\end{equation}
from which we realize that the last two eigenstate are the original two states $\ket{1000}$ and $\ket{0001}$, when $J_1$ is small. Spin transfer can be obtained if three of the levels are in resonance with each other. This can be obtained if $\Delta =\Delta_\pm= 2 (J_2^z \pm J_2^x)$. 

Now we need to consider the remaining subspace $\mathcal{B}_2$ to see if either of the eigenstates here are resonant. Therefore let
\begin{equation}
\ket{ \psi_\pm} = \frac{1}{\sqrt{2}}\left(\left| 01 \right\rangle \pm \left| 10 \right\rangle\right),
\end{equation}
and consider the basis $\{ \ket{0\psi_-1},\allowbreak \ket{0\psi_+1},\allowbreak \ket{0110},\allowbreak \ket{1001},\allowbreak
\ket{1\psi_-0},\allowbreak
\ket{1\psi_+0} \}$, which we transform into using
\begin{equation}\label{eq:coordinateH0}
\mathcal{V} = \left(\begin{smallmatrix}
\frac{1}{\sqrt{2}} & \frac{1}{\sqrt{2}} & 0 & 0 & 0 & 0  \\
\frac{1}{\sqrt{2}} & -\frac{1}{\sqrt{2}} & 0 & 0 & 0 & 0 \\
0 & 0 & 1 & 0 & 0 & 0 \\
0 & 0 & 0 & 1 & 0 & 0 \\
0 & 0 & 0 & 0 &\frac{1}{\sqrt{2}} & \frac{1}{\sqrt{2}} \\
0 & 0 & 0 & 0 & \frac{1}{\sqrt{2}} & -\frac{1}{\sqrt{2}}
\end{smallmatrix}\right),
\end{equation}
which yields the Hamiltonian
\begin{equation}\label{eq:tildeH0}
\tilde{H}_2 = \left(\begin{smallmatrix}
2J_2^x - J_2^z & J_1^z & \sqrt{2}J_1^x & \sqrt{2}J_1^x & 0 & 0 \\
J_1^z & -J_2^z-2J_2^x & -\sqrt{2}J_1^x & -\sqrt{2}J_1^x&   0 & 0  \\
\sqrt{2}J_1^x & -\sqrt{2}J_1^x & - 2J_1^z + J_2^z - \Delta & 0 & \sqrt{2}J_1^x & \sqrt{2}J_1^x \\
\sqrt{2}J_1^x &-\sqrt{2}J_1^x & 0 & - 2J_1^z + J_2^z + \Delta & 
\sqrt{2}J_1^x & \sqrt{2}J_1^x \\
0 & 0 & \sqrt{2}J_1^x & \sqrt{2}J_1^x & 2J_2^x - J_2^z  & J_1^z \\
0 & 0 & \sqrt{2}J_1^x & \sqrt{2}J_1^x & -J_1^z & -J_2^z-2J_2^x
\end{smallmatrix}\right),
\end{equation}
which is approximately diagonal for $J_1 \ll J_2$. Now we need to verify that the desired closed states $\ket{1\psi_\pm0}$ is non-resonant with all the connected other states, and therefore do not evolve. The state has the eigenenergy 
\begin{equation}
\tilde{E}_{1\psi_\pm0} = \pm 2J_2^x - J_2^z.
\end{equation}
Now assume $\Delta = \Delta_+$, then the states $\ket{0110}$ and $\ket{1001}$ obtain the eigenenergies
\begin{align}
\tilde{E}_{0110} =&  - 2J_1^z + J_2^z + \Delta_+ \simeq 3J_2^z + 2J_2^x, \\
\tilde{E}_{1001} =&- 2J_1^z + J_2^z - \Delta_+ \simeq -J_2^z + 2J_2^x,
\end{align} 
which means that $\ket{0\psi_+1}$ is highly non-resonant with all connected states unless $J_2^z=0$. Note that the state $\ket{1\psi_+0}$ have the same energy, but is not directly connected with $\ket{0\psi_+1}$. Note, however that the states $\ket{0\psi_-1}$, $\ket{1\psi_-0}$, and $\ket{1001}$ are resonant.

A similar argument can be made for $\Delta_-$. Assume $\Delta = \Delta_-$, then the states $\ket{0110}$ and $\ket{1001}$ obtain the eigenenergies
\begin{align}
\tilde{E}_{0110} =&  - 2J_1^z + J_2^z + \Delta_- \simeq 3J_2^z - 2J_2^x, \\
\tilde{E}_{1001} =&- 2J_1^z + J_2^z - \Delta_- \simeq -J_2^z - 2J_2^x,
\end{align} 
which means that $\ket{0\psi_-1}$ is highly non-resonant with all connected states unless $J_2^z=0$. Note that the state $\ket{1\psi_-0}$ have the same energy, but is not directly connected with $\ket{0\psi_-1}$. Note, however that the states $\ket{0\psi_+1}$, $\ket{1\psi_+0}$, and $\ket{1001}$ are resonant. This means that the system will oscillate between these three states if started in any one of them. This presents a problem for the state $|1001\rangle$, which we would like to stay unchanged. However, it is not a problem if the period of oscillation between these three state are close to the gate time, since then the system will be in the desired state at the gate time.

Lastly we must also consider the state $|1\psi_\pm 1\rangle$, since we would like it to remain stationary. However, due to the symmetry of the system the subspace which it belongs to, $\mathcal{B}_3$ operates identically to the subspace $\mathcal{B}_1$, and thus the state must behave identically to the states $|0\psi_\pm 0\rangle$, which is indeed stationary as discussed above.

\subsection{Transfer time} \label{app:transferTime}

In order to verify that perfect transfer is achieved and to find the transfer time, we wish to expand the initial and final states in the basis of eigenvectors in the original basis. Therefore we find the eigenvalues of \cref{eq:HmatN4} to be
\begin{subequations}
	\begin{align}
	E_1 = & -J_2^x - \frac{1}{2}\Delta_\pm - \sqrt{4(J_1^x)^2 + \left( \frac{1}{2}\Delta_\pm - J_2^x - J_2^z \right)^2}, \\
	E_2 = & J_2^x - \frac{1}{2}\Delta_\pm - \sqrt{4(J_1^x)^2 + \left( \frac{1}{2}\Delta_\pm + J_2^x - J_2^z \right)^2}, \\
	E_3 = & -J_2^x - \frac{1}{2}\Delta_\pm + \sqrt{4(J_1^x)^2 + \left( \frac{1}{2}\Delta_\pm - J_2^x - J_2^z \right)^2}, \\
	E_4 = & J_2^x - \frac{1}{2}\Delta_\pm + \sqrt{4(J_1^x)^2 + \left( \frac{1}{2}\Delta_\pm + J_2^x - J_2^z \right)^2},
	\end{align}
\end{subequations}
and the corresponding non-normalized eigenvectors in the original basis are
\begin{subequations}\label{eq:EigenBasis}
	\begin{align}
	\ket{\Psi_1} =& \left\lbrace 1, \frac{4(J_1^x)^2 - 2J_2^x(2J_2^z - J_2^z + E_3)}{2J_1^x(J_2^z + E_1)}, \frac{J_2^z - 2J_2^x + E_3}{-2J_1^x}, 1 \right\rbrace, \\
	\ket{\Psi_2} =& \left\lbrace 1, \frac{4(J_1^x)^2 + 2J_2^x(2J_2^z + J_2^z - E_4)}{2J_1^x(J_2^z + E_2)}, \frac{J_2^z - 2J_2^x + E_4}{2J_1^x}, -1 \right\rbrace, \\
	\ket{\Psi_3} =& \left\lbrace 1, \frac{4(J_1^x)^2 - 2J_2^x(2J_2^z - J_2^z + E_1)}{2J_1^x(J_2^z + E_3)}, \frac{J_2^z - 2J_2^x + E_1}{-2J_1^x}, 1 \right\rbrace, \\
	\ket{\Psi_4} =& \left\lbrace 1, \frac{4(J_1^x)^2 + 2J_2^x(2J_2^z + J_2^z - E_2)}{2J_1^x(J_2^z + E_4)}, \frac{J_2^z - 2J_2^x + E_2}{2J_1^x}, -1 \right\rbrace.
	\end{align}
\end{subequations}
We now expand the final and initial state in the basis of the eigenvectors above
\begin{subequations}
	\begin{align}
	\ket{1000} =& \sum_{k =1}^{4}a_k^{(i)} \ket{\Psi_k}, \label{eq:duuuExpanded} \\
	\ket{0001} =& \sum_{k =1}^{4}a_k^{(f)} \ket{\Psi_k}. \label{eq:uuudExpanded}
	\end{align}
\end{subequations}
Since the Hamiltonian of \cref{eq:HmatN4} is a bisymmetric matrix the expansion coefficients are related as $a_k^{(i)} = (-1)^{k}a_k^{(f)}$ \cite{Nield1994}. We thus apply the time evolution operator $U(t) = e^{-iH_1t}$ to the initial state in \cref{eq:duuuExpanded}, and by setting it equal to the final state in \cref{eq:uuudExpanded} we obtain the following condition of perfect state transfer after $t_f$
\begin{equation}
\sum_{k=1}^{4} \left[e^{-iE_kt_f} - (-1)^k\right]a_k^{(i)} \ket{\Psi_k} =0.
\end{equation}
and thus the conditions for perfect state transfer are
\begin{equation}\label{eq:TransReq}
E_kt_f = \begin{cases}
(2n_k + 1)\pi & \text{for } k = 1,3, \\
2n_k\pi & \text{for } k = 2,4 ,
\end{cases}
\end{equation}
thus we find the sufficient conditions for the state $\ket{1000}$ to evolve into the state $\ket{0001}$ to be 
\begin{equation}
|E_{k+1} - E_k|t_f = (2m_k +1)\pi,
\end{equation}
where $m_k$ is an integer since we assume $E_1 < E_2 < E_3 < E_4$. For $\Delta_+$ we find the energy distance between the equidistant levels to be
\begin{align*}
|E_2 - E_1| &= \left|2J_1^x + 2J_2^x - \sqrt{(2J_1^x)^2 + (2J_2^x)^2}\right| \simeq |2J_1^x|, \\
|E_3 - E_2| &= \left|2J_1^x - 2J_2^x + \sqrt{(2J_1^x)^2 + (2J_2^x)^2}\right| \simeq |2J_1^x|, \\
|E_4 - E_3| &= \left|-2J_1^x + 2J_2^x + \sqrt{(2J_1^x)^2 + (2J_2^x)^2}\right| \simeq |4J_2^x +2J_1^x|.
\end{align*}
For $J_1^x \ll J_2^x$ we see the the three lowest levels are equidistant with the spacing $|2J_1^x|$, while the highest energy level is far above the others. Thus we can achieve nearly perfect state transfer for $t = \pi/|2J_1^x|$. A completely similar argument can be made for $\Delta_-$.

Now consider the initial and final states 
\begin{subequations}
	\begin{align}
	\ket{i} &= a\ket{1000} + b\ket{0000}, \\
	\ket{f} &= a\ket{0001} - b\ket{0000},
	\end{align}
\end{subequations}
where $|a|^2 + |b|^2 =1$. The change of sign on the last term is due to the fact that the eigenstate $\ket{0000}$ receives a phase factor of $e^{-i\pi} = -1$ during the transfer, as mentioned in the main text. Once again we expand the states $\ket{\,1000}$ and $\ket{\,0001}$ into the basis of eigenvectors of \cref{eq:EigenBasis}
\begin{subequations}
	\begin{align}
	\ket{i} &= a\sum_{k =1}^{4}a_k^{(i)} \ket{\Psi_k} + b\ket{0000}, \\
	\ket{f} &= a\sum_{k =1}^{4}a_k^{(f)} \ket{\Psi_k} - b\ket{0000},
	\end{align}
\end{subequations}
and once again we time evolve the initial state and set it equal to the final state, yielding the condition for perfect state transfer after time $t_f$
\begin{equation}
a\sum_{k=1}^{4} \left[e^{-iE_kt_f} - (-1)^k\right]a_k^{(i)} \ket{\Psi_k} + b\left[ e^{-iE_- t_f} + 1 \right] \ket{0000}  =0,
\end{equation}
where the eigenenergy of the non-excited state is $E_0 = -\Delta_\pm + J_2^z + 2J_1^z$, which for the case of $\Delta_+$ is $E_0 = -J_2^z - 2J_2^x + 2J_1^z$. Thus besides the original requirements in \cref{eq:TransReq} we also find the requirement 
\begin{equation}
E_0t_f = (2n_0 + 1)\pi,
\end{equation}
where $n_0$ is an integer. Since $\ket{0000}$ is completely unexcited and thus the lowest state we find the condition 
\begin{equation}\label{eq:condition}
|E_1 - E_0|t_f = 2m_0\pi,
\end{equation}
where $m_0$ is a positive integer. We find the distance between the two energy levels
\begin{equation}
|E_1 - E_0| = |2J_1^x + 2J_1^z|.
\end{equation}
Solving for the transfer time in \cref{eq:condition} and choosing $m_0 = 1$ in order to obtain the fastest transfer time we find
\begin{equation}
t_f = \frac{2\pi}{|E_1 - E_0|} = \frac{\pi}{|J_1^x + J_1^z|}.
\end{equation}
From this it is clear that in order to obtain a transfer of the $\ket{0000}$ state in the same time as the states of $\mathcal{B}_{1}$ we must require $J_1^x = J_1^z$.

\section{The $N=5$ case}\label{app:N5}

With five qubits there are still just four interaction coefficients $J_1^{x,z}$ and $J_2^{x,z}$ due to the spatial symmetry, and we still require $J_1^{x,z} \ll J_2^{x,z}$. However, there are now three different qubit frequencies and thus two different detunings, $\Delta \equiv \Delta_2 = \Delta_4$ and $\Delta_3$. According to Ref. \cite{Marchukov2016}, in the basis of $\{ |10000 \rangle,|01000 \rangle,|00100 \rangle,|00010 \rangle,|00001 \rangle \}$ the eigenvalues of the Hamiltonian is
\begin{subequations}
\begin{align}
	E_1 =& \Delta + \frac{1}{2}\Delta_3 + 2 J_2^z, \\
	E_+ \equiv E_2 =& \frac{1}{2}\Delta + J_1^z - J_2^z + \sqrt{8(J_2^x)^2 + \left(\frac{1}{2}(\Delta - \Delta_3) + J_1^z - J_2^z\right)^2},\\
	E_0 = E_3 =& \frac{1}{2}\Delta_3, \\
	E_- \equiv E_4 =& \frac{1}{2}\Delta + J_1^z - J_2^z - \sqrt{8(J_2^x)^2 + \left(\frac{1}{2}(\Delta - \Delta_3) + J_1^z - J_2^z\right)^2,}\\
	E_5 =& \Delta + \frac{1}{2}\Delta_3 + 2 J_2^z,
\end{align}
\end{subequations}
with (non-normalized) corresponding eigenstates
\begin{subequations}
\begin{align}
	|\Psi_1 \rangle =& \{1,0,0,0,0\},\\
	|\Psi_2 \rangle =& \{0,J_2^x,\frac{1}{2}\Delta_3 - E_+,J_2^x,0\},\\
	|\Psi_3 \rangle =& \{0,1,0,-1,0\},\\
	|\Psi_4 \rangle =& \{0,J_2^x,\frac{1}{2}\Delta_3 - E_-,J_2^x,0\},\\
	|\Psi_5 \rangle =& \{0,0,0,0,1\},\\
\end{align}
\end{subequations}
In agreement with the main text we define the states of the middle control qubits as $|1^\pm\rangle_C = \{J_2^x,\frac{1}{2}\Delta_3 - E_\pm,J_2^x\}$ and $|1^0 \rangle_C = \{1,0,-1\} $ in the basis of $\{ |100 \rangle,|010 \rangle,|001\rangle \}$, and we define $|0\rangle_C = \{0,0,0\}$ as the open configuration.

In order to achieve resonant transfer between the states $|10000\rangle $ and $|00001\rangle$ we must find values of the detuning such that on of $E_{\pm,0}$ is equal to $E_1=E_2$. For the case of $E_0 = E_1$ we find $\Delta = 2J_2^z$ and any $\Delta_3$, and the transfer of excitation goes via the resonant state $|0\rangle |1^0\rangle_C |0\rangle$. In the case of $E_1 = E_\pm$ we find that 
\begin{equation}\label{eq:N5config}
	\Delta = -\frac{2(J_2^x)^2}{\frac{1}{2}\Delta_3 + 2J_2^z} + J_2^z,
\end{equation}
where the transfer goes via the states $|0\rangle |1^\pm\rangle_C |0\rangle $ depending on the value of $\frac{1}{2}\Delta_3 + 2J_2^z$; the transfer goes through $|0\rangle |1^+\rangle_C |0\rangle $ for $\frac{1}{2}\Delta_3 + 2J_2^z < 0$ and $|0\rangle |1^-\rangle_C |0\rangle $ for $\frac{1}{2}\Delta_3 + 2J_2^z > 0$.

\begin{figure}
	\centering
	\includegraphics[width=0.5\columnwidth]{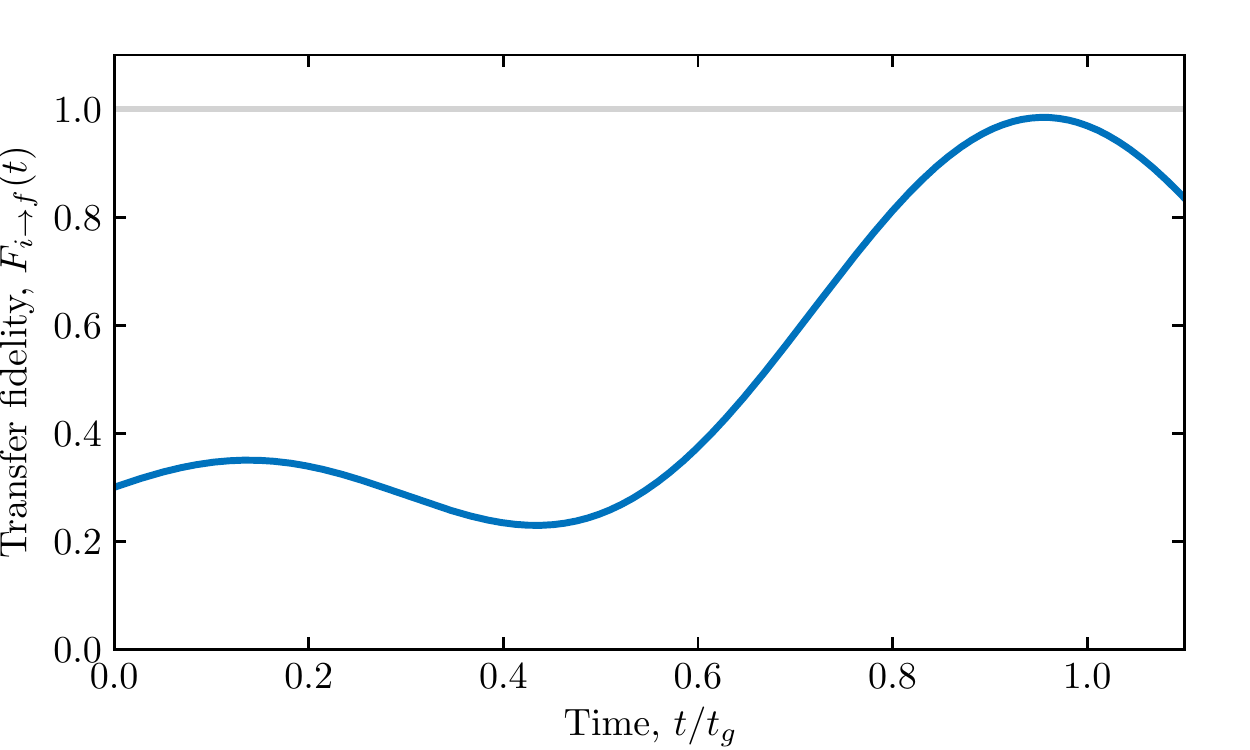}
	\caption{Average fidelity of the system with $N=5$ qubits in the open configuration, i.e. $|0\rangle_C$, with resonant transfer via $|1^0\rangle_C$.}
	\label{fig:N5fidel}
\end{figure}

With these requirements the chain of qubits will work as a transistor as described by Ref. \cite{Marchukov2016}, however it does not guarantee that it will work as a swapping gate in the same manner as for four qubits. This is due to the fact that the transfer time of the states $|10000\rangle$ and $|10001\rangle$ have to coincide as was the case with $N=4$. In fact we find that only the case of $E_1 = E_0$, i.e. resonant transfer via the state $|1^0\rangle_C$, creates the negative swapping gate of Eq. (4) with a fidelity close to 0.99, and operating similar to the case of $N=4$. The remaining two possible configurations of \cref{eq:N5config}, does not create any kind of swapping gate with a fidelity over 0.9.

In the closed configuration of $|1^\pm\rangle_C$, we find that the total excitation remains stationary, as it should, however, we also observe that in the case of a superposition input we acquire an internal phase, depending on the exact configuration of the gate. In the closed configuration of $|1^0\rangle_C$, i.e. when the gate is configured according to \cref{eq:N5config}, we do, however, not pick up such a phase.

\section{In depth analysis of the circuit}\label{app:Indeepth}

\begin{figure}
	\centering
	\includegraphics[width=0.8\columnwidth]{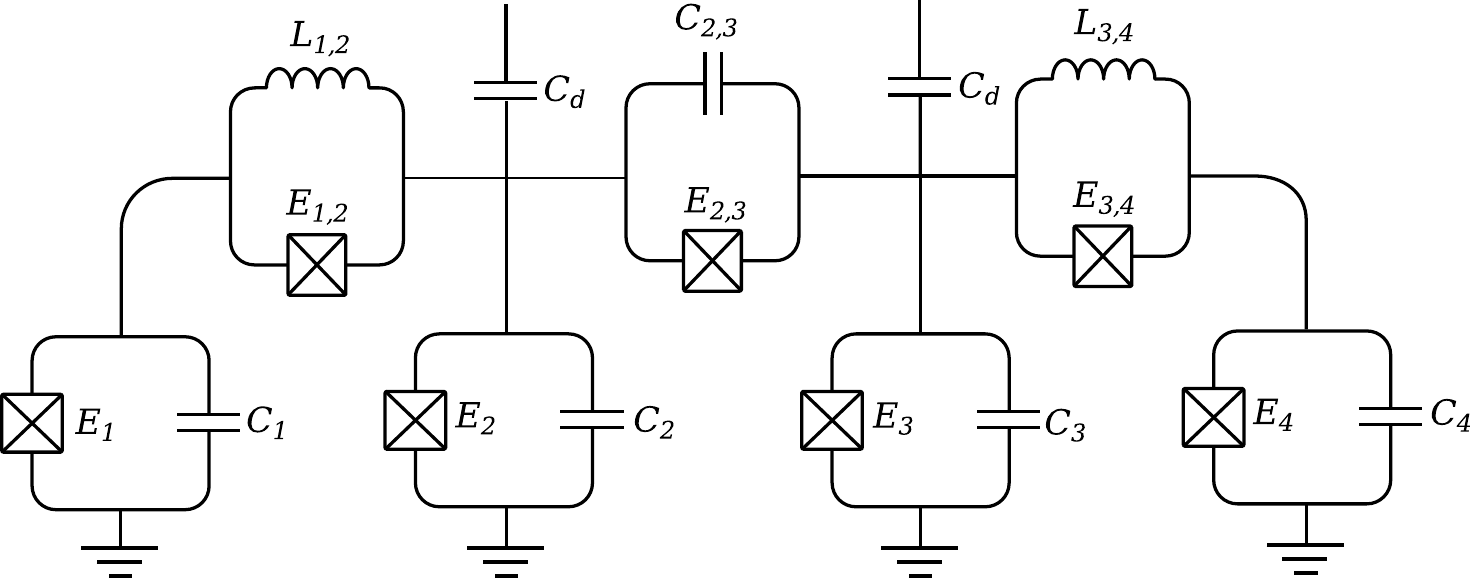}
	\caption{The lumped circuit model. Four grounded Transmon qubits are connected through Josephson junctions and alternating inductors and capacitors.}
	\label{fig:circuit}
\end{figure}

Here follows an in depth derivation of the spin model resulting from the circuit in \cref{fig:circuit} which is equivalent to the circuit in the main text.
The calculations are done for $N=4$, but can easily be expanded to larger $N$, actually it is as simple as expanding the capacitance matrix in \cref{eq:Kmatrix_HN4}. Following the procedure of Refs. \cite{Devoret1997,Devoret2017} we obtain the following Lagrangian
\begin{equation}
\begin{aligned}
L =& 2\sum_{i=1}^{N} C_i \phi_i^2 + 2\sum_{i=1}^{N-1} C_{i,i+1} \left(\dot{\phi}_i-\dot{\phi}_{i+1}\right)^2
+\sum_{i=1}^{N} E_i\cos\phi_i \\ 
&+ \sum_{i=1}^{N-1} E_{i,i+1} \cos\left(\phi_i - \phi_{i+1}\right) - \frac{1}{2}\sum_{i=1}^{N-1} \frac{(2\pi)^2}{L_{i,i+1}} \left(\phi_i - \phi_{i+1}\right)^2,
\end{aligned}
\end{equation}
where the first two terms come from the capacitors and are interpreted as the kinetic terms, and the remaining terms come from the Josephson junctions and inductors and are interpreted as the potential terms. Note that in the present case $C_{1,2} = C_{3,4} = 0$ and $1/L_{2,3} = 0$. Also note that the driving capacitances from the control lines are not in the Lagrangian, but since they are in parallel with the shunting capacitances, $C_i$ they can simply be added to those.
The capacitance matrix becomes
\begin{equation}\label{eq:Kmatrix_HN4}
K = 8\begin{pmatrix}
C_1 & 0 & 0 & 0 \\
0 & C_2 + C_{2,3} & -C_{2,3} & 0 \\
0 & -C_{2,3} & C_3 + C_{2,3} & 0 \\
0 & 0 & 0 & C_4
\end{pmatrix},
\end{equation}
which we note is a block diagonal matrix, hence its inverse matrix must be likewise, which means that the only couplings due to the capacitances are between node 2 and 3. See Eq. (A9) for the actual inverse capacitance matrix With the capacitance matrix we can write the Hamiltonian as
\begin{equation}
H = 4 \vec{p}^T K^{-1} \vec{p} + U(\bm{\phi}),
\end{equation}
where $U(\vec{\phi})$ is the potential due to the Josephson junctions and inductors, which we will now focus on. 

\subsection{Expansion of the potential}

We now do a Taylor expansion of the cosines in the potential around zero. To the desired fourth order we obtain
\begin{equation}
\begin{aligned}
U(\vec{\phi}) =  \sum_{i=1}^{N} \left\lbrace \frac{E_i}{2} \phi_i^2 - \frac{E_i}{24} \phi_i^4 \right\rbrace +  \sum_{i=1}^{N-1} E_{i,i+1}\left[\frac{1}{2} \left(\phi_i - \phi_{i+1}\right)^2 - \frac{1}{24} \left(\phi_i - \phi_{i+1}\right)^4 \right]+ \sum_{i=1}^{N-1} \frac{(2\pi)^2}{2L_{i,i+1}} \left(\phi_i - \phi_{i+1} \right)^2 .
\end{aligned}
\end{equation}
Note that since cosine is an even function, no cubic terms appear in the expansion. 
Expanding the parenthesis yields
\begin{equation*}
\begin{aligned}
U(\vec{\phi}) =&  \sum_{i=1}^{N} \left\lbrace\frac{E_i}{2} \phi_i^2 -  \frac{E_i}{24}\phi_i^4\right\rbrace + \sum_{i=1}^{N-1} \frac{(2\pi)^2}{2L_{i,i+1}} \left(\phi_i^2 + \phi_{i+1}^2  - 2\phi_i\phi_{i+1} \right)\\
&+  \sum_{i=1}^{N-1} \left[\frac{E_{i,i+1}}{2} \left(\phi_i^2 + \phi_{i+1}^2 - 2\phi_i\phi_{i+1} \right)-   \frac{E_{i,i+1}}{24} \left(\phi_i^4 + \phi_{i+1}^4 - 4\phi_i ^3\phi_{i+1} + 6\phi_i ^2\phi_{i+1}^2 - 4\phi_i\phi_{i+1}^3\right)\right].
\end{aligned}
\end{equation*}
Collecting terms we arrive at the full Hamiltonian
\begin{equation}\label{eq:H_EFG}
\begin{aligned}
H =&  \sum_{i=1}^{N} \left[4E_{C,i} p_i^2 +  \frac{1}{2}\left(E_{L,i} + E_{J,i}\right)\phi_i^2 - \frac{E_{J,i}}{24} \phi_i^4 \right] + 8(K^{-1})_{(2,3)} p_2p_3 \\
&+  \sum_{i=1}^{N-1}\left[-\left(E_{i,i+1} + \frac{(2\pi)^2}{L_{i,i+1}}\right) \phi_i\phi_{i+1} + \frac{1}{6} E_{i,i+1} (\phi_i ^3 \phi_{i+1} + \phi_i\phi_{i+1}^3) - \frac{1}{4} E_{i,i+1}\phi_i^2\phi_{i+1}^2 \right],
\end{aligned}
\end{equation}
where the effective energy of the capacitances, $E_{C,i}$, are equal to the corresponding diagonal elements of the inverse of the capacitance matrix. The effective Josephson energies are
\label{eq:EffectiveJ_SCN4}
\begin{equation}
E_{J,i} = E_i + E_{i-1,i} + E_{i,i+1},
\end{equation}
where $E_{0,1} = E_{N-1,N} = 0$. Similarly the effective energies of the inductors are
\label{eq:EffectiveL_SCN4}
\begin{equation}
\begin{aligned}
E_{L,i} =& \frac{(2\pi)^2}{L_{i-1,i}}  + \frac{(2\pi)^2}{L_{i,i+1}},
\end{aligned}
\end{equation}
where $1/L_{0,1} = 1/L_{N-1,N} = 0$. Changing into step operators we obtain
\begin{equation}
\begin{aligned}\label{eq:HspinN4step}
H =&  \sum_{i=1}^{N} \left[S_i b^\dagger_ib_i-E_{J,i}T_i^4 (b^\dagger_i + b_i)^4 \right] + 2(K^{-1})_{(1,2)}(T_2T_3)^{-1} (b^\dagger_2 - b_2)(b^\dagger_3 - b_3) \\
& +  \sum_{i=1}^{N-1}\bigg[-\left(E_{i,i+1} + \frac{(2\pi)^2}{L_{i,i+1}}\right) T_iT_{i+1} (b^\dagger_i + b_i)(b^\dagger_{i+1} + b_{i+1}) - \frac{1}{4} E_{i,i+1}T_i^2T_{i+1}^2(b^\dagger_i + b_i)^2(b^\dagger_{i+1} + b_{i+1})^2 \\
& \qquad\quad +\frac{1}{6} E_{i,i+1} \left\lbrace T_i^3T_{i+1} (b^\dagger_i + b_i)^3(b^\dagger_{i+1} + b_{i+1}) + T_iT_{i+1}^3 (b^\dagger_i + b_i)(b^\dagger_{i+1} + b_{i+1})^3\right\rbrace \bigg],
\end{aligned}
\end{equation}
where we have defined
\begin{subequations}
\begin{align}\label{eq:Tcoeff}
T_n =& \left(\frac{2E_{C,n}}{E_{J,n}+ E_{L,n}}\right)^{1/4},\\
S_n =& 4\sqrt{\frac{1}{2}E_{C,n}\left(E_{L,n} + E_{J,n}\right)}\label{eq:Scoeff}.
\end{align}
\end{subequations}

\subsection{Truncating to a spin model}\label{sec:SpinChain_Truncating}

We are now ready to truncate the Hamiltonian in \cref{eq:H_EFG} into a spin model. The Hamiltonian becomes
\begin{equation}
\begin{aligned}
H =&  -\sum_{i=1}^{N} \left[\frac{1}{2} S_i - \frac{1}{4}E_{J,i}T_i^4\right]\sigma_i^z - 2(K^{-1})_{(2,3)}(T_iT_{i+1})^{-1} \sigma^y_2\sigma^y_3 \\
&+  \sum_{i=1}^{N-1}\left[-\left(E_{i,i+1} + \frac{(2\pi)^2}{L_{i,i+1}}\right) T_iT_{i+1}\sigma^x_i \sigma^x_{i+1} +  \frac{1}{6} E_{i,i+1} (T_i^3T_{i+1} \sigma^x_i \sigma^x_{i+1} + T_iT_{i+1}^3 \sigma^x_i \sigma^x_{i+1})\right. \\
& \qquad - \left.  \frac{1}{4} E_{i,i+1}T_i^2T_{i+1}^2\left(\sigma^z_i \sigma^z_{i+1} - 2(\sigma^z_i + \sigma^z_{i+1})\right) \right],
\end{aligned}
\end{equation}
which can be rewritten more elegantly as
\begin{equation}
\begin{aligned}
H =&  -\frac{1}{2}\sum_{i=1}^{N} \Omega_i \sigma^z_i + 2J_2^y \sigma^y_2\sigma^y_3 + \sum_{i=1}^{N-1}\left[2\tilde{J}_{i,i+1}^x\sigma^x_i \sigma^x_{i+1} + J_{i,i+1}^z\sigma^z_i \sigma^z_{i+1} \right],
\end{aligned}
\end{equation}
with the spin frequencies defined as 
\begin{equation}\label{eq:SpinFreq}
\Omega_i = S_i - \frac{1}{2}E_{J,i}T_i^4 - E_{i-1,i} T_{i-1}^2 T_{i}^2 - E_{i,i+1} T_{i}^2 T_{i+1}^2,
\end{equation}
where $T_0 = T_N = 0$. If we truncate to the three lowest states of the anharmonic oscillator, we find that the energy difference between the first and second excited state is given as
\begin{equation}\label{eq:Omegap}
\Omega_i' = \Omega - \frac{1}{2}E_{J,i}T_i^4.
\end{equation}
Thus the absolute and relative anharmonicity becomes
\begin{equation}
\mathcal{A}_i = \Omega_i' - \Omega_i = -\frac{1}{2}E_{J,i}T_i^4, \qquad \mathcal{A}^r_i = -\frac{1}{2} \frac{E_{J,i}T_i^4}{\Omega_i}.
\end{equation}
The coupling constants are defined as 
\begin{subequations}\label{eq:Js}
	\begin{align}
	\tilde{J}_{i}^x =& -\frac{1}{2}\left(E_{i,i+1} + \frac{(2\pi)^2}{L_{i,i+1}}\right) T_iT_{i+1} + \frac{1}{4}E_{i,i+1}(T_i^3T_{i+1} + T_iT_{i+1}^3) , \\
	J^y_{i}=& - (K^{-1})_{(i,i+1)}(T_iT_{i+1})^{-1},\label{eq:Jy}\\
	J_{i}^z =&-\frac{1}{4}E_{i,i+1}(T_iT_{i+1})^2. \label{eq:Jz}
	\end{align}
\end{subequations}
By transforming into the interaction picture (using \cref{eq:H0}) and defining $J_2^x \equiv \tilde{J}_2^x + J_2^y$ we arrive at the Hamiltonian in Eq.~(3) after doing the rotating wave approximation. We have defined the coefficient $J_2^x$ since the swapping terms $\sigma^x$ and $\sigma^y$ become equivalent after the rotating wave approximation.

From the definition of the coupling constants it is evident that the $J_i^z$ coupling constants are linearly proportional to the Josephson energy. The same is the case for some of the terms in the $J_i^x$ coupling constants, however, the terms related to the inductors or capacitors are not dependent on the Josephson energy, thus making it possible to vary $J_i^x$, without changing $J_i^z$ significantly. The reason we say significantly is due to the fact that $T_n$ in \cref{eq:Tcoeff} depends on the both the Josephson energy, and the inductive and capacitive energies. This dependence is more intricate than the linear dependence shown in \cref{eq:Js}, and therefore an investigation of the parameter space become rather complex. In order to overcome this complexity we employ a numerical algorithm in order to search the parameter space. The result of such a parameter search can be seen in \cref{tab:parameters} at the end of this document.

\section{State preparation driving scheme}\label{app:driving}

In order to operate the gate successively, we need a scheme for preparing the state of the gate. We would like to be able to address the gate exclusively, i.e. opening and closing the gate independently of the left and right qubits. This is possible when the outer qubits are detuned from the gate qubits, i.e. $\Delta$ is sufficiently large, compared to the couplings between the gate qubits and the outer qubits. A large detuning can, in experiments, be obtained by tuning the external fluxes.

We can achieve control of the gate by driving node 2 and 3. This is done by adding capacitors with capacitance $C_d$ to the design of the circuit, connecting the nodes $\phi_2$ and $\phi_3$ to an external field $\varphi_d$ respectively.

The addition of these additional capacitors generates the following extra term in the Lagrangian
\begin{equation}
L_d = \frac{C_d}{2} (\dot{\phi}_2- \dot{\varphi}_{d})^2 + \frac{C_d}{2} (\dot{\phi}_3- \dot{\varphi}_{d})^2.
\end{equation}
We now assume that the external field is given as 
\begin{equation}
\varphi_{d} = \tilde{A}s(t) \sin (\omega t + \theta), \qquad \dot{\varphi}_{d} = \tilde{A}\omega \cos (\omega t + \theta),
\end{equation}
where $s(t)$ is some envelope function, $\tilde{A}$ is the amplitude of the driving, $\omega$ is the driving frequency, and $\theta$ is the phase. We rewrite the driving function as
\begin{align*}
	\varphi_d =& \tilde{A} s(t) \sin(\omega t + \theta)\\
	=& \tilde{A} s(t)\left(\cos\theta \sin\omega t + \sin \theta \cos\omega t\right) \\
	=& \tilde{A}\left( I(t) \sin \omega t + Q(t) \cos \omega t\right),
\end{align*}
where we have adopted the definitions $I(t) = s(t)\cos\theta$ for the in-phase component and $Q(t) = s(t) \sin\theta$ for the out-of-phase component.

Expanding the parenthesis yields
\begin{equation}
L_d = \frac{C_d}{2} \left[ \dot{\phi}_2^2 + \dot{\phi}_3^2 + 2 (\tilde{A}\omega\left( I(t) \cos \omega t - Q(t) \sin \omega t\right))^2 - 2\tilde{A}\omega\left( I(t) \cos \omega t - Q(t) \sin \omega t\right) (\dot{\phi}_2 + \dot{\phi}_3)	\right].
\end{equation}
The first two terms are kinetic terms which can be added to the diagonal of the capacitance matrix, the third term is some irrelevant offset term, while the last term can be used to drive the system. The conjugated momentum is altered slightly
\begin{equation}
\vec{p} = K \dot{\vec{\phi}} + \vec{d},
\end{equation}
where $\vec{d}^T = 2\tilde{A}\omega \cos \omega t (0, 1, 1, 0)$, and thus
\begin{equation}
\dot{\vec{\phi}} = K^{-1} (\vec{p} - \vec{d}).
\end{equation}
This changes the kinetic part of the Hamiltonian into
\begin{align*}
H_\text{kin} =& 4(\vec{p} - \vec{d})^T K^{-1} (\vec{p} - \vec{d}) \\
=& 4 \vec{p}^T K^{-1}\vec{p} + 4\vec{d}^T K^{-1} \vec{d} - 4\vec{p}^T K^{-1} \vec{d} - 4\vec{d}^T K^{-1} \vec{p},
\end{align*}
where the first term is the original kinetic term, the second term is some irrelevant offset while the last two terms are identical driving terms yielding
\begin{equation}
H_d = -8 \dot{\varphi} \left\lbrace \left[  (K^{-1})_{(2,2)} + (K^{-1})_{(3,2)} \right] p_2+ \left[ (K^{-1})_{(2,3)} + (K^{-1})_{(3,3)} \right] p_3 \right\rbrace,
\end{equation}
which can easily be truncated to a spin model
\begin{equation}
H_d = \frac{1}{2}A \left( I(t) \cos \omega t - Q(t) \sin \omega t\right) \left( \sigma^y_2 + \sigma^y_3\right), 
\end{equation}
where
\begin{equation}
A = -8\tilde{A}\omega \left( (K^{-1})_{(2,2)} + (K^{-1})_{(3,2)} \right) T_2^{-1}.
\end{equation}
With this we are now ready to change to the interaction picture using the non-interacting Hamiltonian of \cref{eq:H0}
\begin{align*}
(H_d)_I =& \frac{i}{2} A \left( I(t) \cos \omega t - Q(t) \sin \omega t\right)\left[ (\sigma_2^- + \sigma_3^-) e^{i\Omega_1t} - (\sigma_2^+ + \sigma_3^+) e^{-i\Omega_1t} \right]\\
=&  A \left( I(t) \cos \omega t - Q(t) \sin \omega t\right)\left[ (\sigma_2^y + \sigma_3^y)\cos \Omega_1 t -  (\sigma_2^x + \sigma_3^x) \sin \Omega_1 t \right]\\
=&  \frac{A}{2} \left[ I(t) \left( \cos \delta t  (\sigma_2^y + \sigma_3^y) - \sin \delta t  (\sigma_2^x + \sigma_3^x)\right) + Q(t) \left( -\sin \delta t  (\sigma_2^y + \sigma_3^y) + \cos \delta t  (\sigma_2^x + \sigma_3^x)\right) \right],
\end{align*}
where we have introduced the parameter $ \delta = \omega - \Omega_1$ and we have thrown away all fast rotating terms i.e. terms with $\omega + \Omega_1$. The out-of-phase part of the driving Hamiltonian can be used to minimize leakage to the higher excited states during the driving when the anharmonicity is small \cite{Warren1984,Steffen2003}, using the CRAB driving scheme \cite{Gambetta2011,Motzoi2013} or the GRAPE driving scheme \cite{Rebentrost2009,Khaneja2005,Motzoi2009}.

Like the rest of the Hamiltonian this term preserves the total spin of the two gate qubits, hence it does not mix the singlet and triplet states (note that the spin projection is not preserved, which is why the gate can be driven between the different states). We can therefore ignore the singlet state $\left|1^- \right\rangle_C$, when starting from any of the triplet states shown in Fig. 3. 
The energies difference between the triplet states is found by
\begin{subequations}
	\begin{align}
	\left\langle11\right|H \left|11\right\rangle_C =& \Omega_2 + J_2^z,\\
	\frac{1}{2}\left( \left\langle10\right| + \left\langle01\right| \right) H \left( \left|10\right\rangle_C + \left|01\right\rangle_C \right) =& -J_2^z + 2J_2^x,\\
	\left\langle 00\right| H \left|00\right\rangle_C =& -\Omega_2 + J_2^z,
	\end{align}
\end{subequations}
Rabi oscillations between the closed and open states are then generated by the driving provided the driving frequency matches the energy difference $\omega = |\Omega - 2J_2^z + 2J_2^x|$ and $A\ll J_2^z$.
A $\pi$-pulse, $At = \pi$, would then shift between the $\left| 0 \right\rangle_C$ and $\left| 1^+ \right\rangle_C$ states in in a few microseconds depending on the size of $A$.
Thus using this scheme we can drive between an open and closed gate using merely an external microwave drive.

\section{Including the second excited states of the gate}\label{app:secondEx}

In order to fulfill the requirement of $J_2\gg J_1$ and still keep a short gate time the coupling between the control qubits, $J_2$, must be rather large. This yields the concern of leakage to higher excited states. We therefore need to consider when this becomes a problem. To do this we change the control qubits in to qutrits and investigate the chain in this case. Starting from \cref{eq:HspinN4step} we wish to truncate the middle two qubits into qutrits.

Due to the rather large expression it is advantageous to express part of the Hamiltonian at a time.
Starting with the non-interacting part of the qutrit Hamiltonian in \cref{eq:HspinN4step}, i.e. the terms $i=2,3$ in the first sum, we obtain
\begin{equation}
H_{0,i} \sim \begin{pmatrix}
0 & 0 & -\sqrt{2}E_{J,i}T_i^4/4 \\
0 & S_i-E_{J,i}T_i^4/2& 0 \\
-\sqrt{2}E_{J,i}T_i^4/4 & 0 & 2S_i -3E_{J,i}T_i^4/2
\end{pmatrix},
\end{equation}
where we have subtracted some irrelevant offset term.
Note that the $T$ coefficients are the same for both $i=2$ and 3 due to the symmetry of the circuit. From the matrix representation, we see that there is a coupling between the ground and second excited state. This coupling is unwanted and will, like every other odd powers of couplings, disappear during the rotating wave approximation. For convenience we write the Hamiltonian using braket notation, since spin matrices does not turn out to be a good desirable basis in our case
\begin{equation}
\begin{aligned}
H_{0,i} =& \left(S_i-\frac{1}{2}E_{J,i}T_i^4 \right) \ketbra{1}{1} + \left(2S_i -\frac{3}{2}E_{J,i}T_i^4\right) \ketbra{2}{2}\\
& - \frac{\sqrt{2}}{4}E_{J,i}T_i^4\left(\ketbra{0}{2} + \ketbra{2}{0}\right).
\end{aligned}
\end{equation}
We skip the rest of the non-interacting Hamiltonian for now, since it turns out that there are contributions to the energies of the states from the interacting part of the Hamiltonian. 

For convenience we start by expressing the step-operators in the three level model in braket notation 
\begin{subequations}
	\begin{align}
	b^\dagger_i \pm b_i =& \ketbra{1}{0}_i \pm \ketbra{0}{1}_i + \sqrt{2} \left(\ketbra{2}{1}_i \pm \ketbra{1}{2}_i\right), \\
	\left(\hat{b}_i^\dagger + \hat{b}_i\right)^2 =& \ketbra{0}{0}_i + 3\ketbra{1}{1}_i + 5\ketbra{2}{2}_i + \sqrt{2} \left(\ketbra{2}{0}_i \pm \ketbra{0}{2}_i\right), \\
	\left(\hat{b}_i^\dagger + \hat{b}_i\right)^3 =& \ketbra{1}{0}_i + \ketbra{0}{1}_i + \sqrt{2} \left(\ketbra{2}{0}_i + \ketbra{0}{2}_i\right).
	\end{align}
\end{subequations}
Thus we are ready to consider the target and control qubit $x$-interaction with
\begin{equation}
\begin{aligned}
H^x_{T,C} =K_{T,C}^x\sigma^x_T \left[  \ketbra{1}{0}_C+\ketbra{0}{1}_C + \sqrt{2}  \left(\ketbra{2}{1}_C +\ketbra{1}{2}_C\right)\right]+M_{T,C}^x\sigma_T^x \left[ \ketbra{1}{0}_C +\ketbra{0}{1}_C + 2\sqrt{2} \left(\ketbra{2}{1}_C +\ketbra{1}{2}_C\right)\right],
\end{aligned}
\end{equation}
where we use the notation that as a subscript we have either $T=1,4$ and $C=2,3$ in pairs. We have defined
\begin{subequations}\label{eq:MKxcoeff}
	\begin{align}
	K_{i,j}^x =& -\left(E_{i,j} + \frac{(2\pi)^2}{L_{i,j}}\right) T_iT_{j} + \frac{1}{6} E_{i,j} T_i^3T_{j}, \\
	M_{i,j}^x =& \frac{1}{6} E_{i,j}T_i T_{j}^3.
	\end{align}
\end{subequations}
Note that $2\tilde{J}_1^x = K^x_{T,C} + M_{T,C}^x$, however, due to the factor of 2 on the last term in $\hat{H}_{T,C}^x$, we cannot use $\tilde{J}_1^x$. 
The next term we consider is the $z$-interaction between the target and control qubit
\begin{equation}
H'_{T,C} = J_1^z(2\unitm -\sigma_T^z)\left[  \ketbra{0}{0}_C +  3\ketbra{1}{1}_C + 
5 \ketbra{2}{2}_C +	\sqrt{2}  \left(\ketbra{2}{0}_C +\ketbra{0}{2}_C\right) \right],
\end{equation}
where $J_1^z$ can be found in \cref{eq:Jz}. From this we realize that we obtain not only $z$-interactions, but also corrections to the energies of the qutrits and some terms involving $\ketbra{2}{0}$, which will disappear during the rotating wave approximation, if the conditions are right. Therefore we define
\begin{equation}\label{eq:H12z}
H_{T,C}^z = J_1^z\sigma_T^z\left[  \ketbra{0}{0}_C -  \ketbra{1}{1}_C -
3 \ketbra{2}{2}_C -	\sqrt{2}  \left(\ketbra{2}{0}_C +\ketbra{0}{2}_C\right) \right],
\end{equation}
while we add the contribution to the qutrit energy to the non-interacting Hamiltonian.

\begin{figure}
	\centering
	\includegraphics[width=0.6\columnwidth]{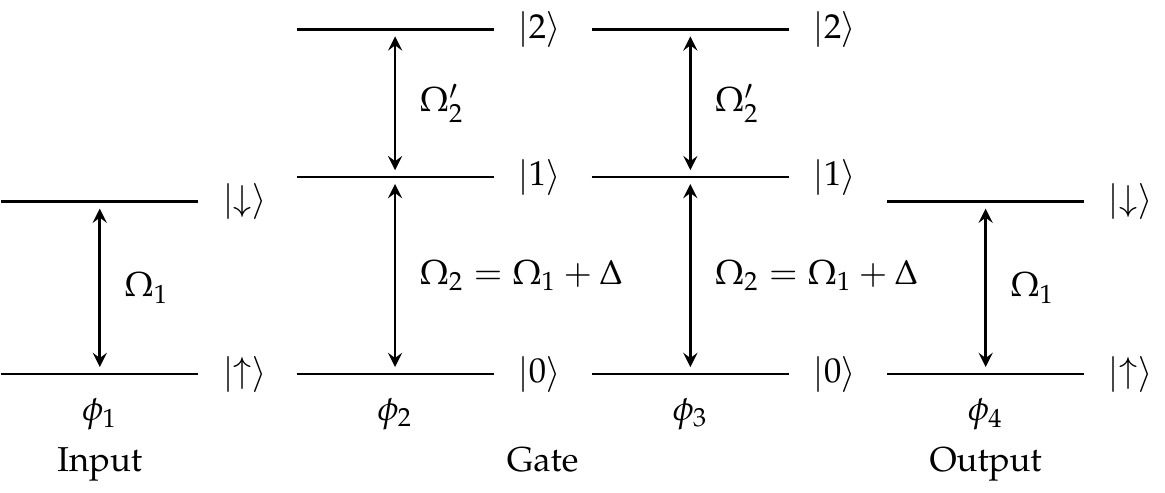}
	\caption{Sketch of the quantum spin gate chain where the third states of the controls have been included, changing the gate into consisting of two qubits. The energy distance between each state has been included in the drawing. The system is the same for the pure qubit gate, just with the two highest levels removed.}
	\label{fig:QutritChain}
\end{figure}

This leaves only the interaction between the two control qutrits. We start with their $y$-interaction
\begin{equation}\label{eq:H23y}
H_{2,3}^y = -2J_2^y \prod_{i=2}^{3} \left[\ketbra{1}{0}_i -\ketbra{0}{1}_i + \sqrt{2} \left(\ketbra{2}{1}_i -\ketbra{1}{2}_i\right)\right] ,
\end{equation}
where $J_2^y$ is defined in \cref{eq:Jy}. Moving on to the $x$-interaction
\begin{equation}\label{eq:H23x}
\begin{aligned}
H^x_{2,3} =& K_{2,3}^x \prod_{i=2}^{3}\left[ \left(\ketbra{1}{0}_i +\ketbra{0}{1}_i\right) + \sqrt{2} \left(\ketbra{2}{1}_i +\ketbra{1}{2}_i\right)\right] \\
&+ M_{2,3}^x \prod_{i=2}^{3}\left[ \left(\ketbra{1}{0}_i +\ketbra{0}{1}_i\right) + 2\sqrt{2} \left(\ketbra{2}{1}_i +\ketbra{1}{2}_i\right)\right].
\end{aligned}
\end{equation}
This leaves the $z$-interaction
\begin{equation}
\begin{aligned}
H'_{2,3} =& J_2^z\prod_{i=2}^{3}\left[  \ketbra{0}{0}_i +  3\ketbra{1}{1}_i + 
5 \ketbra{2}{2}_i +	\sqrt{2}  \left(\ketbra{2}{0}_i +\ketbra{0}{2}_i\right) \right]\\
=&  J_2^z\prod_{i=2}^{3}\left[2 \unitm -\ketbra{0}{0}_i +  \ketbra{1}{1}_i + 
3 \ketbra{2}{2}_i +	\sqrt{2}  \left(\ketbra{2}{0}_i +\ketbra{0}{2}_i\right) \right]\\
=&  -2J_2^z\sum_{i=2}^{3} \left[\ketbra{0}{0}_i - \ketbra{1}{1}_i - 
3 \ketbra{2}{2}_i -	\sqrt{2}  \left(\ketbra{2}{0}_i +\ketbra{0}{2}_i\right)\right] \\
&+ J_2^z\prod_{i=2}^{3}\left[\ketbra{0}{0}_i -\ketbra{1}{1}_i -
3 \ketbra{2}{2}_i -	\sqrt{2}  \left(\ketbra{2}{0}_i +\ketbra{0}{2}_i\right) \right],
\end{aligned}
\end{equation}
where we have thrown away some irrelevant offset term. Once again we find a contribution to the qutrit energy, and some terms related to $\ketbra{2}{0}$, and therefore we only consider the last product as the $z$-interaction calling it $H_{2,3}^z$.
This was the last part of the Hamiltonian and thus the last addition to the energy of the qutrits. Now we can write the full non-interacting Hamiltonian as
\begin{equation}
\begin{aligned}
H_0 =& -\frac{1}{2}\Omega_1(\sigma_1^z + \sigma_4^z) - \frac{1}{2}\Omega_2\left(\ketbra{0}{0}_2-\ketbra{1}{1}_2 + \ketbra{0}{0}_3 - \ketbra{1}{1}_3\right)+ \frac{1}{2}(\Omega_2 + 2\Omega'_2)\left(\ketbra{2}{2}_2 + \ketbra{2}{2}_3\right),
\end{aligned}	
\end{equation}
where $\Omega_i$ can be found in \cref{eq:SpinFreq}. Lastly the energy up to the new state is given in \cref{eq:Omegap} and thus in general we would expect $\Omega_2 \neq \Omega'_2$ due to the anharmonicity. A schematic drawing of the system consisting of an input qubit, two gate qutrits, and an output qubit can be seen in \cref{fig:QutritChain}.

Collecting all terms the full Hamiltonian becomes
\begin{equation}
\begin{aligned}\label{eq:QutritH}
H =& H_0 + \sum_{i=1}^3 (\hat{H}_{i,i+1}^x + \hat{H}_{i,i+1}^z)  + \hat{H}_{2,3}^y+2\sqrt{2} (J_1^z + J_2^z)  \left(\ketbra{2}{0}_2 +\ketbra{0}{2}_2 + \ketbra{2}{0}_3 +\ketbra{0}{2}_3\right).
\end{aligned}
\end{equation}
We now choose our non-interacting Hamiltonian completely equivalent to previously, but with the addition of the second excited state 
\begin{equation}
\begin{aligned}
H_0 =& -\frac{1}{2}\Omega_1(\sigma_1^z + \sigma_4^z) - \frac{1}{2}\Omega_1\left(\ketbra{0}{0}_2-\ketbra{1}{1}_2 + \ketbra{0}{0}_3 - \ketbra{1}{1}_3\right) + \frac{1}{2}(\Omega_2 + 2\Omega'_2)\left(\ketbra{2}{2}_2 + \ketbra{2}{2}_3\right),
\end{aligned}
\end{equation} 
and we wish to perform the rotating wave approximation. All odd powers of exchange operators disappear, as long as the energy difference between the two states is large enough. In our case this is always the case and thus the last term of \cref{eq:QutritH} disappears. Consider now the $x$-interaction $\hat{H}_{T,C}^x$. We write the Pauli $x$-operator as $\sigma_T^x = \ketbra{1}{0}_T + \ketbra{0}{1}_T$ in order to make the notation more obvious. Thus only permutations of $\ketbra{1}{0}_T\ketbra{0}{1}_C$ will survive. This is exactly equivalent to case of the qubits. Similarly permutations of $\ketbra{1}{0}_T\ketbra{1}{2}_C$ can only survive if $\Omega'_2 \sim \Omega_1$. This leaves
\begin{equation}
\begin{aligned}
(H^x_{T,C})_I =& J_1^x  \left(\ketbra{0}{1}_T\ketbra{1}{0}_C+\ketbra{1}{0}_T\ketbra{0}{1}_C\right) \\ &+ \sqrt{2} (K_{T,C}^x + 2M_{T,C}^x) \left(\ketbra{0}{1}_T\ketbra{2}{1}_Ce^{i(\Omega_1 - \Omega'_2)t} +\ketbra{1}{0}_T\ketbra{1}{2}_Ce^{-i(\Omega_1 - \Omega'_2)t}\right).
\end{aligned} 
\end{equation}
However, if $|\Omega_1-\Omega'_2| \gg |\sqrt{2} (K_{T,C}^x + 2M_{T,C}^x)|$ the last two terms will rotate away as well. As $|\sqrt{2} (K_{T,C}^x + 2M_{T,C}^x)|$ is usually the close to the size of the coupling $J_1^x$ when we configure the circuit as a the swapping gate these term will almost always rotate away. We are left with 
\begin{equation}
(H^x_{T,C})_I = J_1^x  \left(\ketbra{0}{1}_T\ketbra{1}{0}_C+\ketbra{1}{0}_T\ketbra{0}{1}_C\right),
\end{equation}
and the interaction resembles the original interaction for when the chain was entirely qubits. 

Turning to $\hat{H}_{T,C}^z$ we see that only the last two terms containing $\ketbra{2}{0}_C$ and $\ketbra{0}{2}_C$ obtain a phase factor of $e^{i(\Omega_2+\Omega'_2)t}$, which makes the terms rotate away, and thus the Hamiltonian becomes
\begin{equation}\label{eq:H12zI}
(H_{T,C}^z)_I = J_1^z\sigma_T^z\left(  \ketbra{0}{0}_C -  \ketbra{1}{1}_C - 
3 \ketbra{2}{2}_C \right).
\end{equation}
The last part of the Hamiltonian is the interaction between the gate qutrits. Starting from the $x$- and $y$-interaction, which we can deal with together since only a sign differs, we realize that the terms $\ketbra{0}{1}_i$ receives a phase of $e^{i\Omega_2t}$, while terms on the form $\ketbra{1}{2}_i$ receive a phase of $e^{i\Omega_2't}$. Thus taking the products in \cref{eq:H23y,eq:H23x} we realize that all phases with a sum of $\Omega_2'$ and/or $\Omega_2$ rotate away rapidly, while terms with differences are kept. Thus focusing on \cref{eq:H23y} we obtain
\begin{equation}\label{eq:H23yI}
\begin{aligned}
(H_{2,3}^y)_I =& 2J_2^y\Big[\ketbra{1}{0}_2\ketbra{0}{1}_3 + \ketbra{1}{0}_2\ketbra{0}{1}_3 +2 \left(\ketbra{2}{1}_2\ketbra{1}{2}_3 +\ketbra{1}{2}_2\ketbra{2}{1}_3\right) \\
& + \sqrt{2}\left( \left\lbrace\ketbra{2}{1}_2\ketbra{0}{1}_3 + \ketbra{0}{1}_2\ketbra{2}{1}_3\right\rbrace e^{i(\Omega_2-\Omega'_2)t}+\left\lbrace \ketbra{1}{2}_2\ketbra{1}{0}_3 + \ketbra{1}{0}_2\ketbra{1}{2}_3\right\rbrace e^{-i(\Omega_2-\Omega'_2)t} \right)\Big].
\end{aligned}
\end{equation}
The products in \cref{eq:H23x} are handled identically yielding a total $x$- and $y$-interaction term

\begin{equation}
\begin{aligned}
(H_{2,3}^{xy})_I =& 2J_2^x\left(\ketbra{1}{0}_2\ketbra{0}{1}_3 + \ketbra{1}{0}_2\ketbra{0}{1}_3\right) +4R_{2,3}^x \left(\ketbra{2}{1}_2\ketbra{1}{2}_3 +\ketbra{1}{2}_2\ketbra{2}{1}_3\right) \\
& + 2\sqrt{2}P_{2,3}^x\left(\left\lbrace\ketbra{2}{1}_2\ketbra{0}{1}_3 + \ketbra{0}{1}_2\ketbra{2}{1}_3\right\rbrace e^{i(\Omega_2-\Omega'_2)t} +\left\lbrace \ketbra{1}{2}_2\ketbra{1}{0}_3 + \ketbra{1}{0}_2\ketbra{1}{2}_3\right\rbrace e^{-i(\Omega_2-\Omega'_2)t}  \right),
\end{aligned}
\end{equation}
where we have set $R_{2,3}^x = J_2^y + K_{2,3}^x + 4M_{2,3}^x$ and $P_{2,3}^x = J_2^y + K_{2,3}^x + 2M_{2,3}^x$. 
Thus if the anharmonicity is large enough such that $|\Omega_2-\Omega'_2|\gg 2\sqrt{2}|J_2^y + K_{2,3}^x + 2M_{2,3}^x|$ the last four terms will rotate away and the third level will be completely decoupled from the two lowest states. However, in our case we require the coupling between the two control qubits/qutrits to be strong and thus the anharmonicity will not be large enough to rotate these terms away. However, we do not expect this to be a problem since the swapping is only inside the gate, and a closed gate never consist of enough excitation to reach any state higher than the first excited state.

Lastly we have the $z$-interaction between the gate qutrits. This is handled equivalently to \cref{eq:H12zI} and after the fast rotating terms have been removed we obtain
\begin{equation}
\begin{aligned}
(H_{2,3}^z)_I = J_2^z\left(  \ketbra{0}{0}_2 -  \ketbra{1}{1}_2 - 
3 \ketbra{2}{2}_2 \right)\left(  \ketbra{0}{0}_3 -  \ketbra{1}{1}_3 - 
3 \ketbra{2}{2}_3 \right) + 2  J_2^z\left(\ketbra{2}{0}_2\ketbra{0}{2}_3 +\ketbra{0}{2}_2\ketbra{2}{0}_3\right),
\end{aligned}
\end{equation}
where the second term is a swap-term between a double excitation of the two qutrits, and will probably not make much difference since our troubles starts when any of the two qutrits become double excited, and thus a swap of the doubly excited qutrits is irrelevant for us.

Combining the above parts of the full Hamiltonian in the interaction picture we obtain
\begin{equation}\label{eq:fullQutritH}
\begin{aligned}
H = &-\frac{1}{2}\Delta\left(\ketbra{0}{0}_2-\ketbra{1}{1}_2 + \ketbra{0}{0}_3 - \ketbra{1}{1}_3\right) +  2J_1^x  \left(\ketbra{0}{1}_1\ketbra{1}{0}_2+\ketbra{1}{0}_1\ketbra{0}{1}_2\right)\\
& + J_1^z\left( \ketbra{0}{0}_1 - \ketbra{1}{1}_1 \right)\left(  \ketbra{0}{0}_2 -  \ketbra{1}{1}_2 - 
3 \ketbra{2}{2}_2 \right) \\
&+ J_2^z\left(  \ketbra{0}{0}_2 -  \ketbra{1}{1}_2 - 
3 \ketbra{2}{2}_2 \right)\left(  \ketbra{0}{0}_3 -  \ketbra{1}{1}_3 - 
3 \ketbra{2}{2}_3 \right)\\
& + 2  J_2^z\left(\ketbra{2}{0}_2\ketbra{0}{2}_3 +\ketbra{0}{2}_2\ketbra{2}{0}_3\right)+ 2J_2^x\left(\ketbra{0}{1}_2\ketbra{1}{0}_3 + \ketbra{1}{0}_2\ketbra{0}{1}_3\right) \\
&+4R_{2,3}^x \left(\ketbra{2}{1}_2\ketbra{1}{2}_3 +\ketbra{1}{2}_2\ketbra{2}{1}_3\right) \\
& + 2\sqrt{2}P_{2,3}^x\left(\left\lbrace\ketbra{2}{1}_2\ketbra{0}{1}_3 + \ketbra{0}{1}_2\ketbra{2}{1}_3\right\rbrace e^{i(\Omega_2-\Omega'_2)t} +\left\lbrace \ketbra{1}{2}_2\ketbra{1}{0}_3 + \ketbra{1}{0}_2\ketbra{1}{2}_3\right\rbrace e^{-i(\Omega_2-\Omega'_2)t}  \right)\\
&+  2J_{3,4}^x  \left(\ketbra{0}{1}_4\ketbra{1}{0}_3+\ketbra{1}{0}_4\ketbra{0}{1}_3\right) + J_{3,4}^z\left( \ketbra{0}{0}_4 - \ketbra{1}{1}_4 \right)\left(  \ketbra{0}{0}_3 -  \ketbra{1}{1}_3 - 
3 \ketbra{2}{2}_3 \right).
\end{aligned}
\end{equation}
Despite this intricate Hamiltonian, we realize that it is still excitation preserving, and we can thus consider each subspace, $\mathcal{B}_k$, where $k = 0,\dots,6$, individually. By comparing the new three state Hamiltonian in \cref{eq:fullQutritH}, with the old spin Hamiltonian in Eq.~(3), we see that the systems behaves identically within these subspaces. 

The only remaining subspace we need to consider are $\mathcal{B}_2$, which is the 
same as in \cref{app:Indeepth} with the addition of the two states $\ket{0200}$ and $\ket{0020}$. Thus we only need to calculate the part of the Hamiltonian concerning these two states, as the rest is calculated in \cref{eq:H0}. This gives us the block matrix 
\begin{equation}
H_2 = \left(\begin{array}{c|c}
H_0 & W^\dagger \\
\hline
W & V
\end{array}\right),
\end{equation}
where
\begin{subequations}
	\begin{align}
	V =& \begin{pmatrix}
	\Delta/2-3J_2^z-3J_1^z & 2J_2^z \\
	2J_2^z & \Delta/2-3J_2^z-3J_1^z \\
	\end{pmatrix}, \\
	W =& 2\sqrt{2} P_{2,3}^xe^{i(\Omega_2-\Omega_2')t}\begin{pmatrix}
	0 & 0 & 1 & 0 & 0 & 0 \\
	0 & 0 & 1 & 0 & 0 & 0 \\
	\end{pmatrix},
	\end{align}
\end{subequations}
and $H_0$ is given in \cref{eq:HmatN4}. From this we see that the addition of the two new states only interfere with the state $\ket{0110}$, thus we can perform our coordinate transformation of \cref{eq:coordinateH0} to $H_0$ as before, without affecting any of the new states. The conclusion to this is that the states $\ket{0\psi_\pm1}$ and $\ket{1\psi_\pm0}$ are still approximate eigenstates and thus stationary, however, $\ket{0110}$ is no longer an eigenstate. Thus we still have a closed gate, as long as the remaining three eigenstates are not in resonance with the closed state. We do not calculate these eigenstates here, but it is sufficient to say that they are not in resonance.

Due to inclusion of the third states the resonance between the states $|1001\rangle$ and $|1\psi_\mp0\rangle$ and $|0\psi_\mp1\rangle$ only hold approximately meaning that the oscillation between these three states is slowed down.

\clearpage
\thispagestyle{empty}
\begin{sidewaystable}
	\centering
	\caption{Circuit and corresponding spin model parameters. The parameters are found by minimizing a cost function which returns a low value when the spin parameters obey the requirement from the main text. Due to the rather large parameter space (8 circuit parameters), there are several solutions to this, and some (but far from all) are shown in the table below. Above the single line shows the case of $\Delta_+ = 2(J_2^z + J_2^x)$, while below the line shows the case $\Delta_- = 2(J_2^z - J_2^x)$. These numbers are shown in the column $\Delta_\pm$ and should be compared with the actual detuning, $\Omega_2-\Omega_1$, which is not shown since these are identical for all cases. The circuit parameters can be seen in \cref{fig:circuit}. The colored lines (number 6 and 11) corresponds to the parameters used in the simulation shown in Fig. 4 of the main text.}
	\label{tab:parameters}
	\begin{ruledtabular}
	\begin{tabular}{=c|+c+c+c+c+c+c+c+c|+c+c+c+c+c+c+c|+c+c+c+c}
	\multicolumn{1}{c}{} & \multicolumn{8}{c}{Circuit parameters} & \multicolumn{7}{c}{Spin model parameters} & \multicolumn{4}{c}{Second excited state} \\
	& $E_1$ & $E_2$ & $E_{1,2}$ & $E_{2,3}$ & $C_1$ & $C_2$ & $C_{2,3}$ & $L_{1,2}$ & $\Omega_1$ & $\Omega_2$ & $J_1^x$ & $J_1^z$ & $J_2^x$ & $J_2^z$ & $\Delta_\pm$ &  $\mathcal{A}_1^r$ & $\mathcal{A}_2^r$ & $K_{2,3}^x$ & $M_{2,3}^x$ \\
	\# & [$2\pi$GHz] & [$2\pi$GHz] & [$2\pi$GHz] & [$2\pi$GHz] & [fF] & [fF] &[fF] & [nH] & [$2\pi$GHz] & [$2\pi$GHz] & [$2\pi$MHz] & [$2\pi$MHz] & [$2\pi$MHz] & [$2\pi$MHz] & [$2\pi$MHz] & [\%] & [\%] & [$2\pi$MHz] & [$2\pi$MHz] \\
	\toprule 
	1 & $ 532.0 $ & $ 464.1 $ & $ 187.9 $ & $ 410.4 $ & $ 729.4 $ & $ 65.6 $ & $ 279.4 $ & $ 36.7 $ & $ 10.7 $ & $ 11.3 $ & $ 42.0 $ & $ 42.0 $ & $ -494.6 $ & $ 804.1 $ & $ 619.0 $ &  $ 0.16 $ & $ 4.66 $ & $ 2680.3 $ & $ -536.1 $ \\
	2 & $ 119.2 $ & $ 179.6 $ & $ 86.8 $ & $ 165.5 $ & $ 956.6 $ & $ 332.9 $ & $ 440.2 $ & $ 81.2 $ & $  4.4 $ & $  3.8 $ & $ 43.2 $ & $ 43.2 $ & $ -764.3 $ & $ 453.7 $ & $ -621.1 $ &  $ 0.13 $ & $ 10.48 $ & $ 1512.5 $ & $ -302.5 $ \\
	3 & $ 215.4 $ & $ 219.4 $ & $ 100.5 $ & $ 200.6 $ & $ 907.9 $ & $ 223.6 $ & $ 23.8 $ & $ 70.9 $ & $  6.1 $ & $  6.7 $ & $ 47.3 $ & $ 46.9 $ & $ -552.1 $ & $ 844.6 $ & $ 585.0 $ &  $ 0.19 $ & $ 10.28 $ & $ 2815.4 $ & $ -563.1 $ \\
	4 & $ 310.2 $ & $ 371.2 $ & $ 74.9 $ & $ 272.4 $ & $ 183.5 $ & $ 25.8 $ & $ 649.2 $ & $ 93.9 $ & $ 16.1 $ & $ 19.2 $ & $ 44.9 $ & $ 44.9 $ & $ 962.1 $ & $ 562.8 $ & $ 3049.7 $ &  $ 0.51 $ & $ 0.52 $ & $ 1876.0 $ & $ -375.2 $ \\
	5 & $ 690.3 $ & $ 486.3 $ & $ 169.0 $ & $ 393.8 $ & $ 466.5 $ & $ 48.7 $ & $ 391.9 $ & $ 40.4 $ & $ 15.1 $ & $ 14.2 $ & $ 33.3 $ & $ 33.2 $ & $ -987.4 $ & $ 498.0 $ & $ -978.9 $ &  $ 0.21 $ & $ 1.37 $ & $ 1660.0 $ & $ -332.0 $ \\ 
	\rowstyle{\color{blue}}6 & $ 561.6 $ & $ 438.5 $ & $ 186.0 $ & $ 397.1 $ & $ 926.3 $ & $ 76.2 $ & $ 240.4 $ & $ 37.3 $ & $  9.7 $ & $ 10.7 $ & $ 40.9 $ & $ 40.9 $ & $ -540.4 $ & $ 1007.1 $ & $ 933.4 $ &  $ 0.15 $ & $ 6.88 $ & $ 3357.1 $ & $ -671.4 $ \\
	7 & $ 379.3 $ & $ 254.1 $ & $ 90.3 $ & $ 220.9 $ & $ 850.3 $ & $ 48.2 $ & $ 84.0 $ & $ 79.1 $ & $  8.3 $ & $ 12.1 $ & $ 35.5 $ & $ 35.5 $ & $ 764.2 $ & $ 1107.6 $ & $ 3743.7 $ &  $ 0.21 $ & $ 4.74 $ & $ 3692.0 $ & $ -738.4 $ \\
	8 & $ 699.7 $ & $ 601.5 $ & $ 230.9 $ & $ 517.1 $ & $ 664.2 $ & $ 62.5 $ & $ 507.3 $ & $ 29.4 $ & $ 12.8 $ & $ 12.6 $ & $ 35.7 $ & $ 35.7 $ & $ -696.6 $ & $ 582.7 $ & $ -227.7 $ &  $ 0.16 $ & $ 2.63 $ & $ 1942.3 $ & $ -388.5 $\\
	\hline
	9 & $ 156.9 $ & $ 192.3 $ & $ 80.7 $ & $ 181.9 $ & $ 950.5 $ & $ 734.8 $ & $ 839.6 $ & $ 89.9 $ & $  5.1 $ & $  5.8 $ & $ 51.7 $ & $ 51.7 $ & $ 727.6 $ & $ 1081.2 $ & $ 707.2 $  & $ 0.21 $ & $ 14.39 $ & $ 3604.1 $ & $ -720.8 $ \\
	10 & $ 699.8 $ & $ 611.2 $ & $ 236.7 $ & $ 547.7 $ & $ 676.7 $ & $ 64.1 $ & $ 185.7 $ & $ 29.0 $ & $ 12.7 $ & $ 12.9 $ & $ 45.4 $ & $ 45.4 $ & $ 857.1 $ & $ 965.2 $ & $ 216.1 $  & $ 0.15 $ & $ 4.73 $ & $ 3217.3 $ & $ -643.5 $ \\ 
	\rowstyle{\color{blue}} 11 & $ 313.1 $ & $ 199.6 $ & $ 78.3 $ & $ 185.2 $ & $ 994.4 $ & $ 152.6 $ & $ 302.3 $ & $ 92.1 $ & $  7.0 $ & $  7.4 $ & $ 34.7 $ & $ 34.7 $ & $ 936.2 $ & $ 1137.7 $ & $ 403.0 $  & $ 0.21 $ & $ 10.57 $ & $ 3792.2 $ & $ -758.4 $ \\
	12 & $ 144.6 $ & $ 177.9 $ & $ 72.0 $ & $ 167.2 $ & $ 965.4 $ & $ 619.3 $ & $ 20.0 $ & $ 98.8 $ & $  4.8 $ & $  4.2 $ & $ 38.1 $ & $ 38.1 $ & $ 980.3 $ & $ 641.8 $ & $ -677.1 $  & $ 0.22 $ & $ 11.29 $ & $ 2139.2 $ & $ -427.8 $ \\
	13 & $ 113.3 $ & $ 277.9 $ & $ 118.7 $ & $ 264.2 $ & $ 987.7 $ & $ 706.6 $ & $ 583.2 $ & $ 58.8 $ & $  4.3 $ & $  3.6 $ & $ 51.9 $ & $ 51.9 $ & $ 852.9 $ & $ 538.2 $ & $ -629.5 $  & $ 0.02 $ & $ 11.81 $ & $ 1793.9 $ & $ -358.8 $ \\
	14 & $ 167.1 $ & $ 251.7 $ & $ 104.2 $ & $ 238.1 $ & $ 978.3 $ & $ 381.5 $ & $ 999.8 $ & $ 67.6 $ & $  5.2 $ & $  4.7 $ & $ 45.4 $ & $ 45.4 $ & $ 965.8 $ & $ 724.8 $ & $ -482.0 $  & $ 0.15 $ & $ 11.73 $ & $ 2415.9 $ & $ -483.2 $ \\
	15 & $ 178.4 $ & $ 352.0 $ & $ 144.2 $ & $ 324.9 $ & $ 953.8 $ & $ 230.0 $ & $ 488.4 $ & $ 47.7 $ & $  5.4 $ & $  5.0 $ & $ 41.4 $ & $ 41.4 $ & $ 661.6 $ & $ 446.5 $ & $ -430.4 $  & $ 0.08 $ & $ 6.46 $ & $ 1488.2 $ & $ -297.6 $ \\
	16 & $ 132.9 $ & $ 272.4 $ & $ 112.1 $ & $ 255.5 $ & $ 970.5 $ & $ 418.1 $ & $ 408.4 $ & $ 61.9 $ & $  4.6 $ & $  3.7 $ & $ 41.5 $ & $ 41.5 $ & $ 882.6 $ & $ 438.3 $ & $ -888.5 $  & $ 0.07 $ & $ 8.73 $ & $ 1461.1 $ & $ -292.2 $
	\end{tabular}
	\end{ruledtabular}
\end{sidewaystable}